\documentclass[twocolumn]{aastex63}
\usepackage{multirow,amsmath,amsthm,mathtools,mathrsfs}
\usepackage{chngcntr,amssymb,bbold}
\usepackage{natbib}
\usepackage{color,xcolor}

\usepackage{bm}
\usepackage{hyperref} 

\def	\cm		{{\rm {cm}}}
\def	\K		{{\rm K}}
\def	\g		{{\rm {g}}}
\def	\mum	{{\mu \rm{m}}}

\def \bea {\begin{eqnarray}}
\def \ena {\end{eqnarray}}   
\def \Td {T \rm{_d}}

\def \NHt {N(\rm H_2)}
\def \S {\mathcal{S}}

\def	\Angstrom	{\,{\rm \AA}}		

\def	\B	{{\rm B}}

\def	\C	{{\rm C}}

\def	\cm	{\,{\rm cm}}
\def	\d	{{\rm d}}

\def	\D	{{\rm D}}

\def	\erg	{\,{\rm erg}}

\def	\g	{\,{\rm g}}
\def	\gas	{\,{\rm gas}}
\def	\G	{{\rm G}}

\def	\H	{{\rm H}}

\def	\N	{{\rm N}}
\def	\nH	{n_{\rm H}}

\def	\pc	{\,{\rm pc}}

\def	\s	{\,{\rm s}}

\def	\V	{{\rm V}}

\def    \Bv     	{\bf  B}

\def    \gas     	{{\rm gas}}

\accepted{}

\shorttitle{BALLAD-POL: II - Grain alignment physics}

\graphicspath{{./}{figures/}}

\begin{document}

\title{\LARGE{\textbf{B-fields And dust in interstelLar fiLAments using Dust POLarization (BALLAD-POL): II. Testing the Radiative Torque Paradigm in Musca and OMC-1}}}

\author[0000-0002-5913-5554]{Nguyen Bich Ngoc}
\affiliation{Department of Astrophysics, Vietnam National Space Center, Vietnam Academy of Science and Technology, 18 Hoang Quoc Viet, Hanoi, Vietnam, \href{mailto:capi37capi@gmail.com}{capi37capi@gmail.com}} 
\affiliation{Graduate University of Science and Technology, Vietnam Academy of Science and Technology, 18 Hoang Quoc Viet, Hanoi, Vietnam}

\author[0000-0003-2017-0982]{Thiem Hoang}
\affiliation{Korea Astronomy and Space Science Institute, 776 Daedeokdae-ro, Yuseong-gu, Daejeon 34055, Republic of Korea, \href{mailto:thiemhoang@kasi.re.kr}{thiemhoang@kasi.re.kr}}
\affiliation{University of Science and Technology, Korea, 217 Gajeong-ro, Yuseong-gu, Daejeon 34113, Republic of Korea}

\author[0000-0002-2808-0888]{Pham Ngoc Diep}
\affiliation{Department of Astrophysics, Vietnam National Space Center, Vietnam Academy of Science and Technology, 18 Hoang Quoc Viet, Hanoi, Vietnam, \href{mailto:capi37capi@gmail.com}{capi37capi@gmail.com}}
\affiliation{Graduate University of Science and Technology, Vietnam Academy of Science and Technology, 18 Hoang Quoc Viet, Hanoi, Vietnam}

\author[0000-0002-6488-8227]{Le Ngoc Tram}
\affiliation{Max-Planck-Institut f\"{u}r Radioastronomie, Auf dem H\"{u}gel 69, 53-121, Bonn, Germany}

\begin{abstract}

Polarization of starlight and thermal dust emission caused by aligned dust grains is a valuable tool to characterize magnetic fields (B-fields) and constrain dust properties. However, the grain alignment physics is not fully understood. To test the RAdiative Torque (RAT) paradigm, including RAT Alignment (RAT-A) and Disruption (RAT-D), we use dust polarization observed by {\it Planck} and SOFIA/HAWC+ toward two filaments, Musca and OMC-1, with contrasting physical conditions. Musca, a quiescent filament, is ideal for testing RAT-A, while OMC-1, an active star-forming region OMC-1, is most suitable for testing RAT-D. We found that polarization fraction, $P$, decreases with increasing polarization angle dispersion function, $\S$, and increasing column density, $\NHt$, consistent with RAT-A. However, $P$ increases with increasing dust temperature, $\Td$, but decreases when $\Td$ reaches a certain high value. We compute the polarization fraction for the ideal models with B-fields in the plane of sky based on the RAT paradigm, accounting for depolarization effect by B-field tangling. We then compare the realistic polarization model with observations. For Musca with well-ordered B-fields, our numerical model successfully reproduces the decline of $P$ toward the filament spine (aka. polarization hole), having higher $\NHt$ and lower $\Td$, indicating the loss of grain alignment efficiency due to RAT-A. For OMC-1, with stronger B-field variations and higher temperatures, our model can reproduce the observed $P-\Td$ and $P-N(\rm H_{2})$ relations only if the B-field tangling and RAT-D effect are incorporated. Our results provide more robust observational evidence for the RAT paradigm, particularly the recently discovered RAT-D.

\end{abstract}


\section{Introduction}\label{sec:intro}

Dust is a crucial element of the interstellar medium (ISM) and has a significant impact on a variety of astrophysical processes, such as gas heating and cooling, the formation of stars and planets, and surface astrochemistry (see \citealt{draine2003erratum} for an overview). Moreover, the polarization of starlight \citep{Hall.1949, Hiltner.1949} or polarized thermal dust emission \citep{Hildebrand.1988} induced by aligned dust grains are widely used to observe magnetic fields (B-fields), which are predicted to regulate the formation and evolution of molecular clouds and filaments, star formation, and stellar feedback via outflows/jets (see \citealt{crutcher2021review, pattle2019a,pattle2022} for reviews). Thermal dust polarization has also been used to map B-fields in nearby galaxies (e.g., \citealt{Lopez-Rodriguez2022}) and even in high-z galaxies \citep{Geach2023Natur}. Moreover, dust polarization is a valuable tool to probe dust physics and dust properties (size, shape, and composition, see e.g., \citealt{Draine2021ApJ}).

Dust polarimetry is based on the assumption that dust grains are systematically aligned with the ambient B-fields with their longest axes perpendicular to the fields. Hence, the polarization vectors of thermal dust emission are perpendicular to the B-fields projected on the plane of the sky (POS). Therefore, by rotating the observed polarization angle by 90$^\circ$, we can infer the morphology of the B-fields (e.g., \citealt{Hildebrand.1988}).

To establish dust polarimetry as a reliable diagnostic tool to probe B-fields and dust properties, it is necessary to fully understand the detailed physics of grain alignment. The leading mechanism of grain alignment is based on the Radiative Torques (RATs), arising from the differential scattering and extinction of an anisotropic radiation field by non-spherical grains \citep{Dolginov.1976,draine1997radiative,lazarianhoang2007}. The key prediction of the Radiative Torque Alignment (RAT-A) theory is that the degree of grain alignment depends on the local conditions, including the gas density and radiation, such that the alignment degree decreases with increasing the local density mainly due to gas randomization but increases with the radiation field intensity or dust temperature due to the more efficient alignment of grains in stronger radiation fields \citep{hoang2014grain,lazarian2019magnetic}. As a result, the RAT-A theory predicts two key features, including (1) the anti-correlation of the polarization degree, $P$, with the gas density, and (2) a correlation between $P$ and the intensity of the radiation field (or dust temperature) (see \citealt{hoang2021polhole}). 

Previous tests of RAT-A using starlight polarization support the key predictions of the RAT theory (see \citealt{andersson_2015}). For instance, observations toward different regions usually report the general trend of the decrease in the polarization fraction with the gas column density, which is known as the polarization hole, by both optical-NIR (e.g., \citealt{whittet2008efficiency}) and far-IR/submillimeter polarization observations \citep{ward2000first}. Numerical modeling by \citet{whittet2008efficiency} and \citet{hoang2021polhole} showed that the polarization hole could be explained by the loss of grain alignment toward high-density and low radiation field regions based on the RAT-A theory. For a comprehensive review of the successful tests for the RAT-A theory using starlight polarization, please refer to a review by \cite{andersson_2015}.

Nevertheless, to date, there is still a lack of comprehensive testing of the RAT-A mechanism using thermal dust polarization. Specifically, there is an ongoing debate regarding the exact cause of the polarization hole observed in far-IR/sub-mm toward molecular clouds and star-forming regions.
For the diffuse and translucent clouds, it is found that the B-field fluctuations are likely the main cause of the polarization hole, instead of grain alignment loss predicted by the RAT-A, as found from both observations (e.g., \citealt{planck19}) and numerical simulations (e.g., \citealt{Seifried2019SILCCZoom}). Furthermore, the inclination angle of the B-fields with respect to the LOS significantly affects the polarization fraction due to the projection effect \citep{HoangTruong23}. In particular, observations revealed that $P$ does not always increase with $\Td$ \citep{Planck2020p12}, which is opposite to the key prediction of the RAT-A theory. 


\citet{hoang2019ratdnatastro} proposed a new mechanism called RAdiative Torque Disruption (RAT-D), which is expected to resolve the later puzzle. The basics of the RAT-D mechanism is that large grains illuminated by an intense radiation field are spun up to suprathermal rotation by RATs, the same driver of grain alignment. When the centrifugal stress induced by grain rotation exceeds the grain's tensile strength, large grains will be disrupted into smaller ones. Numerical modeling of thermal dust polarization by \citet{lee_2020} using a simplified (DustPOL-py) code based on the RAT-A and RAT-D mechanisms (the so-called RAT paradigm) showed that the depletion of large grains by RAT-D causes a decrease in the polarization fraction at long wavelengths when the dust temperature is sufficiently large, which successfully reproduces the anticorrelation seen by {\it Planck}. The basic idea of this effect is the following. For a fixed maximum grain size ($a_{\rm max}$), a higher radiation strength or dust temperature can align smaller grains, broadening the grain size distribution aligned and increasing the thermal dust polarization fraction. However, when RAT-D happens, large grains are disrupted, removing grains from $a_{\rm max}$ to the disruption size ($a_{\rm disr}$), which results in a narrower grain size distribution and decreases the polarization fraction (see Figure 7 in \citealt{Tram2022} for a detail).
To test the RAT paradigm, previous studies \citep{tram2021ophiuchi, tram2021doradus, thuong22, ngoc2021observations} analyzed far-IR and submillimeter polarization data observed toward star-forming regions by the High-resolution Airborne Wideband Camera Plus (HAWC+) instrument \citep{harper2018hawc} on board the Stratosphere Observatory for Infrared Astronomy (SOFIA) and James Clark Maxwell Telescope (JCMT). They found that $P$ tends to increase with $\Td$ but then decreases when $T_{\rm d}$ exceeds some high temperatures, supporting the RAT paradigm. Later, detailed numerical modeling by \citet{tram2021ophiuchi, tram2021doradus} using the DustPOL-py  code successfully reproduced the $P - \Td$ trend observed in the OMC-1 and $\rho$ Ophiuchus A molecular clouds, which showed the initial success of the joint effect of RAT-A and RAT-D. 

However, the effects of the B-field fluctuations on the polarization hole and the variation of $P-T_{\rm d}$ have not been considered yet in the previous tests of the RAT paradigm using polarization data (see \citealt{Tram2022} for a review). It is, therefore, necessary to take into account the effects of B-field fluctuations to achieve a complete testing of the RAT paradigm.
 
Filaments are ubiquitous and form a complex network within most interstellar clouds, as revealed by {\it Herschel}. Filaments are considered the earliest stage of the star formation process in which the gas density and dust temperature vary significantly across the filaments (e.g., \citealt{andre2014filamentary}), which makes them ideal targets for testing the RAT paradigm. Therefore, the two main objectives of our study are to (1) test the RAT paradigm and (2) study the effect of the B-field fluctuations on the net thermal dust polarization of aligned dust in filaments. To achieve these objectives, we use dust polarization data observed toward Musca by {\it Planck} and OMC-1 by SOFIA/HAWC+ and perform synthetic modeling of the dust polarization with our DustPOL-py  code. Our selection of these filaments is intended to account for the different local properties, e.g., the gas density and radiation field (with or without embedded protostars). While Musca is one of the simplest filaments without embedded protostars, Orion is the closest high-mass star-forming region. With low dust temperatures and small B-field tangling, Musca is a perfect target to test RAT-A. With high dust temperatures, OMC-1 is the ideal target for testing RAT-D. We use our DustPOL-py code to perform pixel-by-pixel polarization modeling of the observed polarization data with the key local physical parameters inferred from observations.

This paper is the second one in our series aimed at characterizing the properties of B-fields And dust in interstelLar fiLAments using Dust POLarization (BALLAD-POL). Our first paper focuses on the B-fields and dust in a massive filament, G11.11-0.12 \citep{Ngoc23}. We structure this paper as follows. In Section \ref{sec:obs}, we describe the observational data on which we work, and our data analysis techniques are described in Section \ref{sec:Analysis}. Section \ref{sec:model} recalls the basis of our model and shows the comparison with the observational data. We discuss the results in Section \ref{sec:discuss}, and summarize the main findings in Section \ref{sec:conclusions}.

\section{Observations}
\label{sec:obs}
\subsection{Polarization Data}
\label{subsec:poldata}

\begin{table*}[!htb]
\begin{centering}
\caption{Properties of Musca and OMC-1 and their observations.}\label{tab1}
\begin{tabular}{c c c c c c c}
\hline
\hline
Objects& Distance& Facility & Wavelength & Beam size & Physical scale\footnote{The scale of the beam size}& Reference \\
\hline
Musca  & $170$ (pc)& {\it Planck}     & $850$ ($\mu m$)&$5 \arcmin$ & $0.24$ (pc)& \cite{Planck2020p12}\\
OMC-1& $388$ (pc)& SOFIA& $214$ ($\mu m$)&$18\arcsec.2$ & $0.03$ (pc)& \cite{chuss2019hawc+}\\
\hline \hline
\end{tabular}
\end{centering}
\end{table*}

We use the thermal dust polarization data observed by {\it Planck} and SOFIA/HAWC+ from Musca and OMC-1, respectively. Information on these filaments and their observations is provided in Table \ref{tab1}.

For linear polarization, the polarization states are defined by the Stokes parameters $I$, $Q$, and $U$. Because of the presence of noise in $Q$ and $U$, the polarized emission intensity, $PI$, is debiased by \citep{Vaillancourt2006}:
\begin{eqnarray}
PI=\sqrt{Q^2+U^2-0.5(\delta Q^2 + \delta U^2)},
\end{eqnarray} 
where $\delta Q$ and $\delta U$ are the uncertainties on $Q$ and $U$, respectively. Assuming that $\delta Q$ and $\delta U$ are uncorrelated, the uncertainty on $PI$ is \citep{Gordon2018}
\begin{eqnarray}
\delta PI=\sqrt{\frac{Q^2\delta Q^2+U^2\delta U^2}{Q^2+U^2}}.
\end{eqnarray}

The polarization degree, $P$, and its uncertainty, $\delta P$, are calculated as 
\begin{eqnarray}
P (\%)=100\times\frac{PI}{I},
\end{eqnarray}
and
\begin{eqnarray}
\delta P(\%)=100\times \sqrt{\frac{\delta PI^2}{I^2}+\frac{\delta I^2(Q^2+U^2)}{I^4}} {\rm ,} 
\end{eqnarray}
where $\delta I$ is the uncertainty on $I$ which is provided together with $I$ from observations.

The polarization angle, $\theta$, and its uncertainty, $\delta \theta$, are calculated as follows:
\begin{eqnarray}
\theta=\frac{1}{2}\mathrm{atan2}\left(\frac{U}{Q}\right)
\end{eqnarray}
\begin{eqnarray}
\delta \theta=0.5\times\frac{\sqrt{U^2\delta Q^2+Q^2\delta U^2}}{(Q^2+U^2)}.
\end{eqnarray}

For the {\it Planck} data, $\theta$ is calculated by $\theta=\frac{1}{2}\mathrm{atan2}(-U/Q)$ because of the different convention used for the Stokes parameters \citep{Planck2020p12}. 

The polarization direction of thermal dust emission is perpendicular to the projection of B-fields in the POS; therefore, the B-field orientation angles on the POS are obtained by rotating $90^\circ$ from the polarization angle $\theta$. Following the IAU convention, the angles of the B-field lines are east of north, ranging from 0$^\circ$ to 180$^\circ$.

\subsection{Polarization Angle Dispersion Function}
\label{subsec:S}

To quantify the tangling of B-fields at small scales, we compute the polarization angle dispersion function, denoted as $\mathcal{S}$ (see Section 3.3 in \citealt{Planck2020p12}). For each pixel at location $\bm{x}$, $\mathcal{S}(\bm{x}, \delta)$ is determined as the root mean square of the polarization angle difference between the reference pixel $\bm{x}$ and pixel $i$, $\mathcal{S}_{xi}=\theta(\bm{x})-\theta(\bm{x}+{\delta})$. Pixel $i$ is situated on a circle with $\bm{x}$ as the center and a radius of $\delta$:

\begin{equation}
\label{eq:ang_dis_func}
 \mathcal{\S}^2(\bm{x},\delta) = \frac{1}{N}
\sum _{i=1}^{N}\mathcal{S}_{xi}^2,
\end{equation}
with $N$ is the number of pixels lying on the circle. For the current study, we calculate $\mathcal{S}(\bm{x},\delta)$ for $\delta$ taking a value of one beam size of the observations.
Due to noise on the Stokes parameters $Q$ and $U$, $\S$ is biased. The variance of angle dispersion function, $\sigma_\S$, and the debiased $\S_{\rm db}$ for a certain pixel $\bm{x}$ are calculated as (see Section 3.5 of \citealt{Planck2020p12}):

\begin{equation}\label{eq:sigma_s}
\begin{split}
\sigma^2_\S& = \frac{\delta\theta^2(\bm{x})}{N^2\S^2} \left(
\sum _{i=1}^{N}\mathcal{S}_{xi} \right)^2 \\
& + \frac{1}{N^2\S^2}\sum _{i=1}^{N}(\mathcal{S}_{xi})^2\delta\theta^2(\bm{x}+\bm{\delta}),
\end{split}
\end{equation}
and
\begin{equation}
\S_{\rm db}^2 = \S^2 - \sigma^2_\S.
\end{equation}

Here, $\S_{\rm db}$ can be positive or negative depending on whether the true value is smaller or larger than the random polarization angle of $52^{\circ}$ \citep{alina2016polarization, serkowski1962}. The data points with $\S \le \sigma_\S$ are removed. Hereafter, we refer to $\S_{\rm db}$ as $\S$ for convenience.

\subsection{Dust Temperature and Column Density Maps}
\label{subsec:tdnadnh2maps}
For further studies, we adopt the dust temperature and column density maps of Musca and OMC-1 from previous studies. These maps are obtained by fitting a modified black body function to the spectral energy distributions (SED).

\begin{figure*}[!htb]
\centering
\includegraphics[trim=5.3cm .5cm 5.3cm 1.cm,clip,width=6.cm]{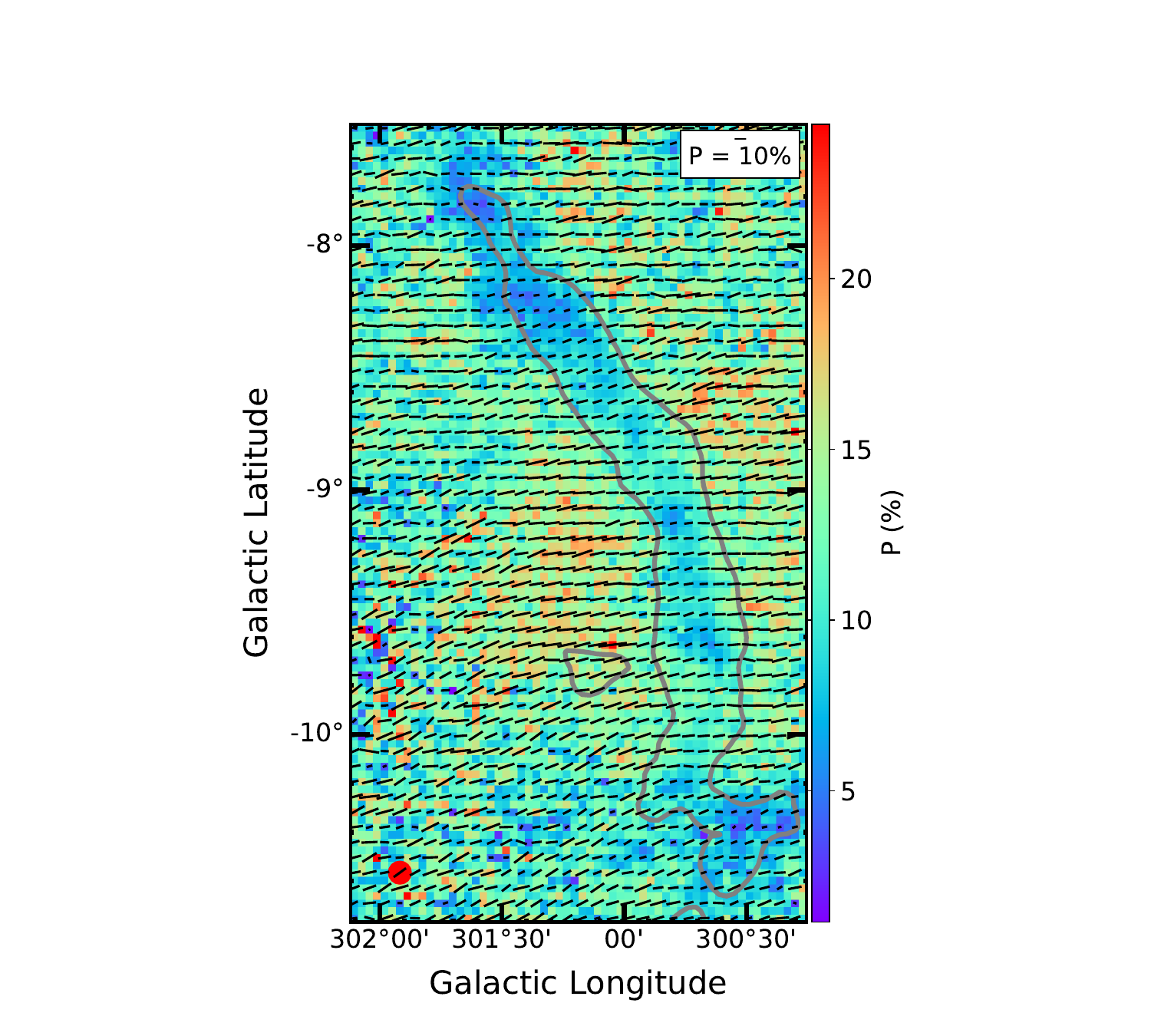}
\includegraphics[trim=6.3cm .5cm 5.3cm 1.cm,clip,width=5.6cm]{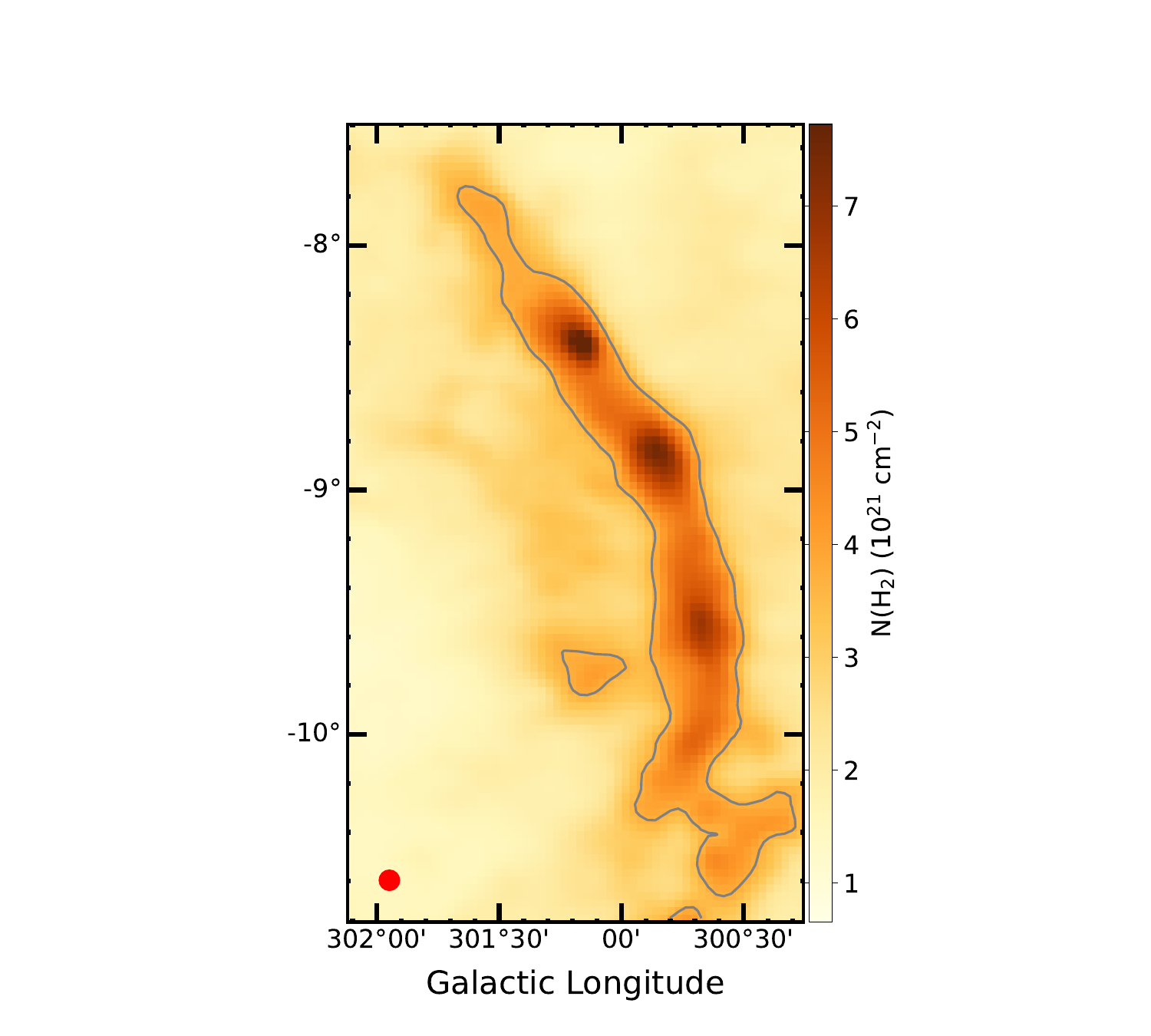}
\includegraphics[trim=6.3cm .5cm 5.3cm 1.cm,clip,width=5.6cm]{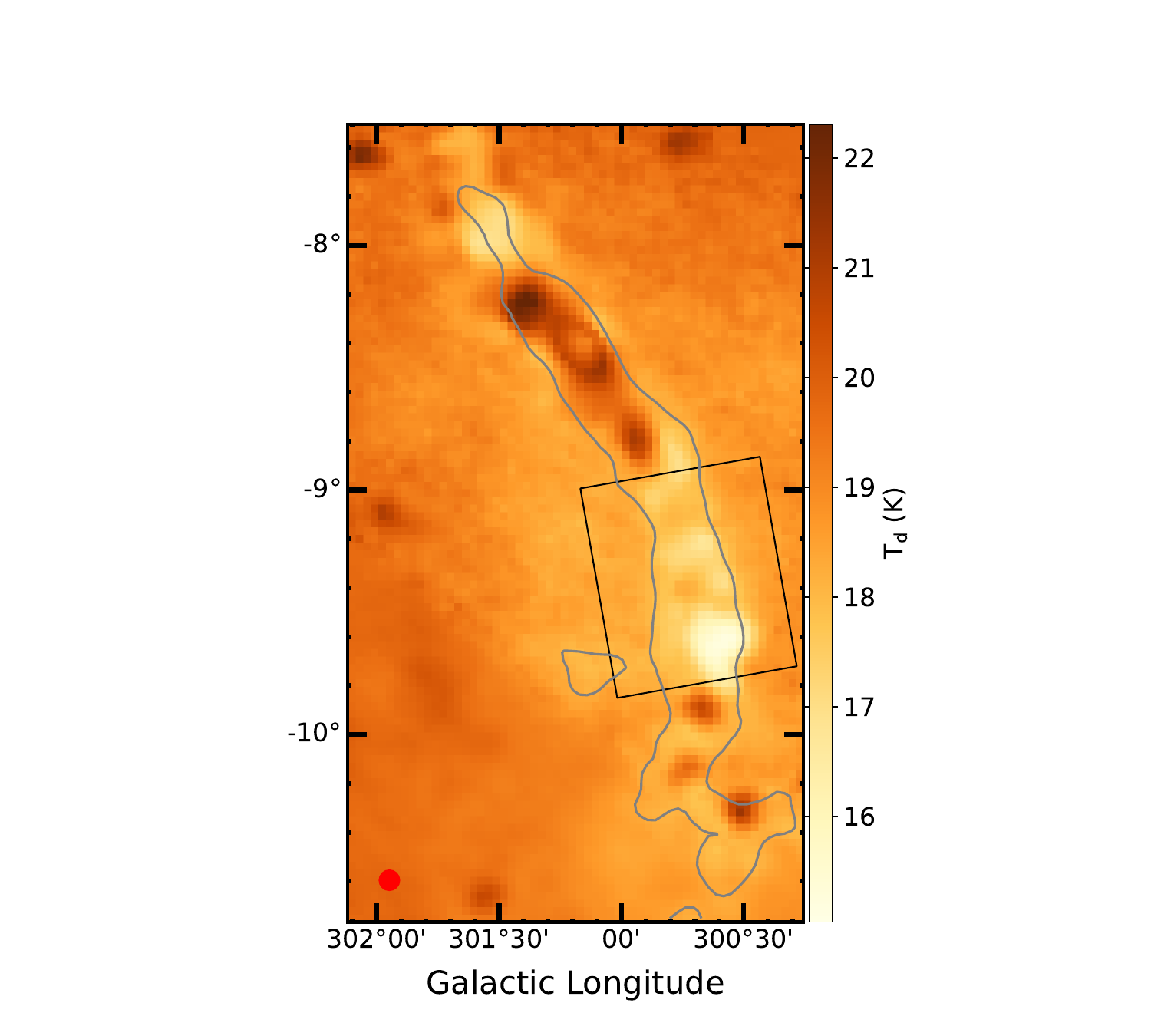}
\caption{Musca observed by {\it Planck}. Left: B-field map. The length of the line segment is proportional to the polarization fraction $P$ with a 10\% reference. Center: Column density. Right: Dust temperature. The half-power beam width (HPBW) of the maps $5\arcmin$ is shown by the red circle in the lower left corner. The gray contours are $3.7 \times 10^{21} \cm^{-2}$. The black rectangle enclosing the southern section of the filament is the area of interest for further analysis.}\label{fig:musca_maps} 
\end{figure*}

\textbf{Musca}: The Planck Collaboration uses data collected by {\it Planck} at 850, 550, 350 $\mum$ and IRAS at 100 $\mum$ to fit with a modified black-body function to obtain all-sky maps of dust temperature, optical depth, and spectral index \citep{Planck_11_mbb}. The resolution of the resulting maps is $5\arcmin$. The dust temperature and column density maps of Musca shown in Figure \ref{fig:musca_maps} (center and right) are extracted from these {\it Planck} all-sky maps.

\textbf{OMC-1}: The dust temperature and column density maps of OMC-1 are adopted from \citet{chuss2019hawc+}. These two maps were obtained by fitting SED from emission in the range of 53 $\mum$ to 35 mm, including data from SOFIA at 53, 89, 154, and 214 $\mum$; {\it Herschel} at 70, 100, 160, and 250 $\mum$; James Clerk Maxwell Telescope (JCMT) at 850 $\mum$. The longest wavelength of the combined Green Bank Telescope and Very Large Array data at 3.5 and 35 mm was used to subtract the free–free emission caused by UV in the region. The final temperature and density maps have a resolution of $18.\arcsec2$. Those maps are shown in Figure \ref{fig:orionmaps} (center and right).

\section{Data Analysis}
\label{sec:Analysis}

To investigate the physics of dust alignment, the analysis of polarization fraction and total intensity is frequently employed for simplification. However, total intensity is a function of column density and dust temperature, making it challenging to disentangle the effects of these two parameters from the polarization effect. Therefore, this study examines the relationship between the polarization fraction, $P$, gas column density, $\NHt$, and dust temperature, $\Td$. Additionally, we analyze the variation of the angle dispersion function, $\S$, to explore the impact of B-field tangling on dust polarization.

In the following subsections, we will first introduce Musca and OMC-1. Secondly, we will study the dependence of $P$ on physical conditions, e.g., column density and dust temperature. This will reveal the basic properties of grain alignment and disruption in the regions. Then, we will present the dependence of $P$ on the polarization angle dispersion function $\mathcal{S}$ to investigate the effects of the B-field tangling.

\subsection{Musca}
\label{subsec:obsmusca}

\begin{figure*}[!htb]
\centering
\includegraphics[trim=0.5cm 0cm 1.5cm 1cm,clip,width=0.47\textwidth]{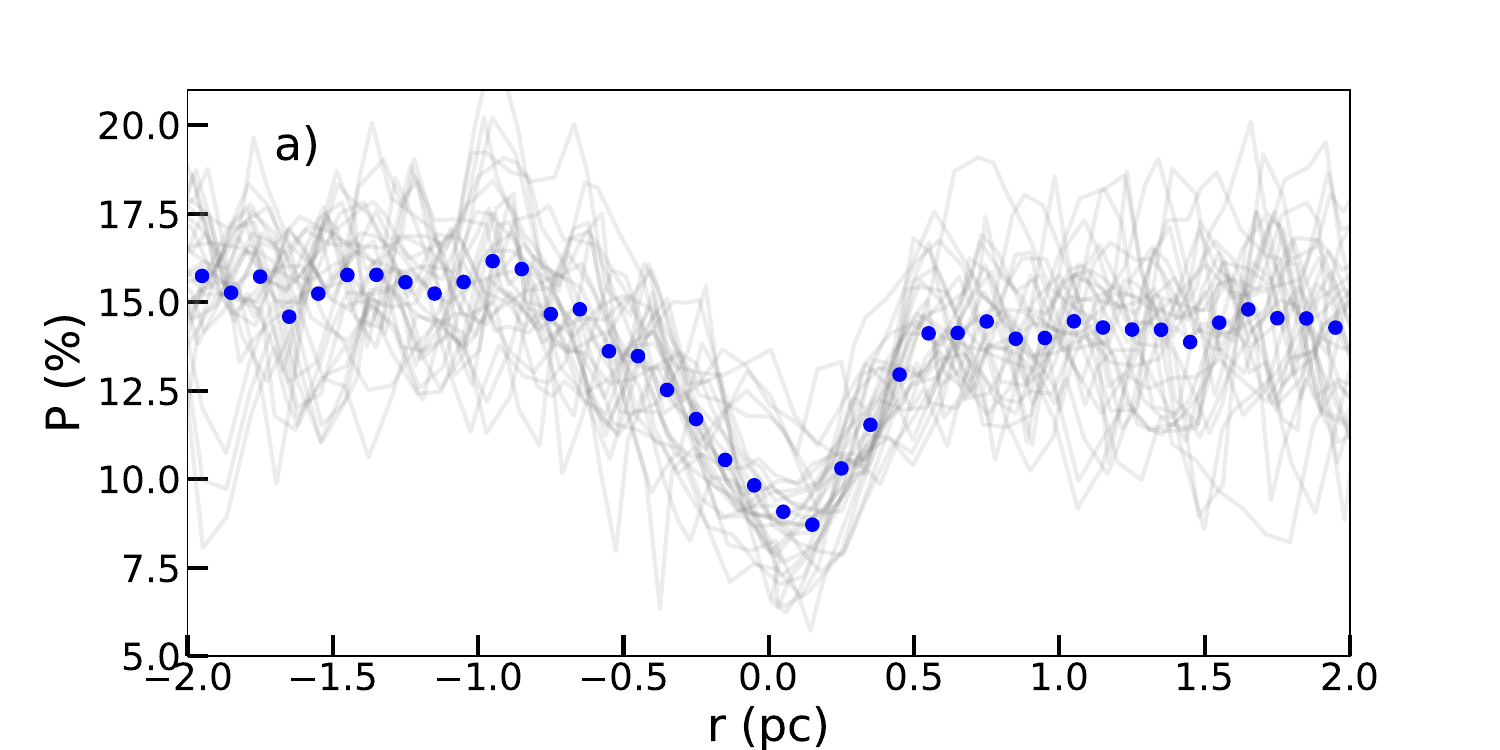}
\includegraphics[trim=0.5cm 0cm 1.5cm 1cm,clip,width=0.47\textwidth]{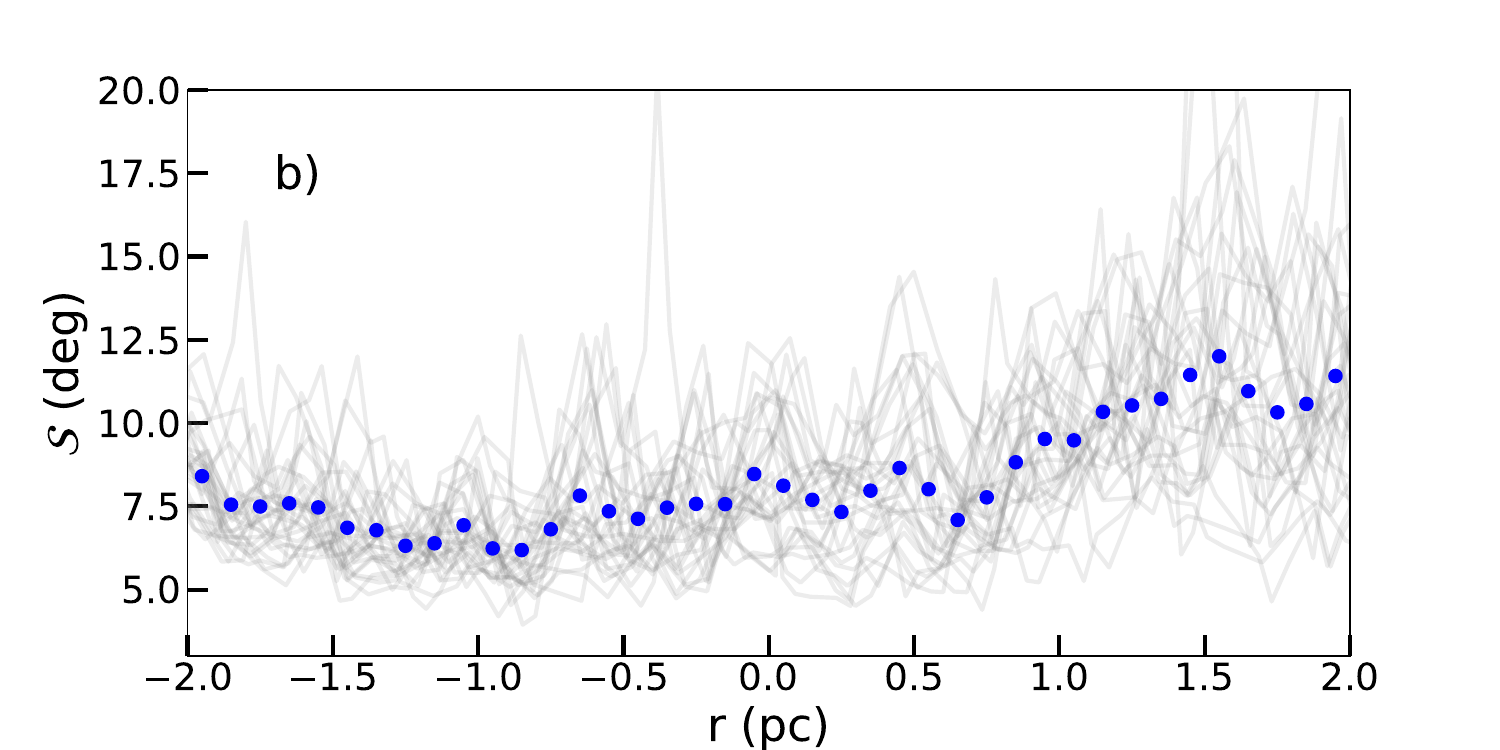}
\includegraphics[trim=0.5cm 0cm 1.5cm 1cm,clip,width=0.47\textwidth]{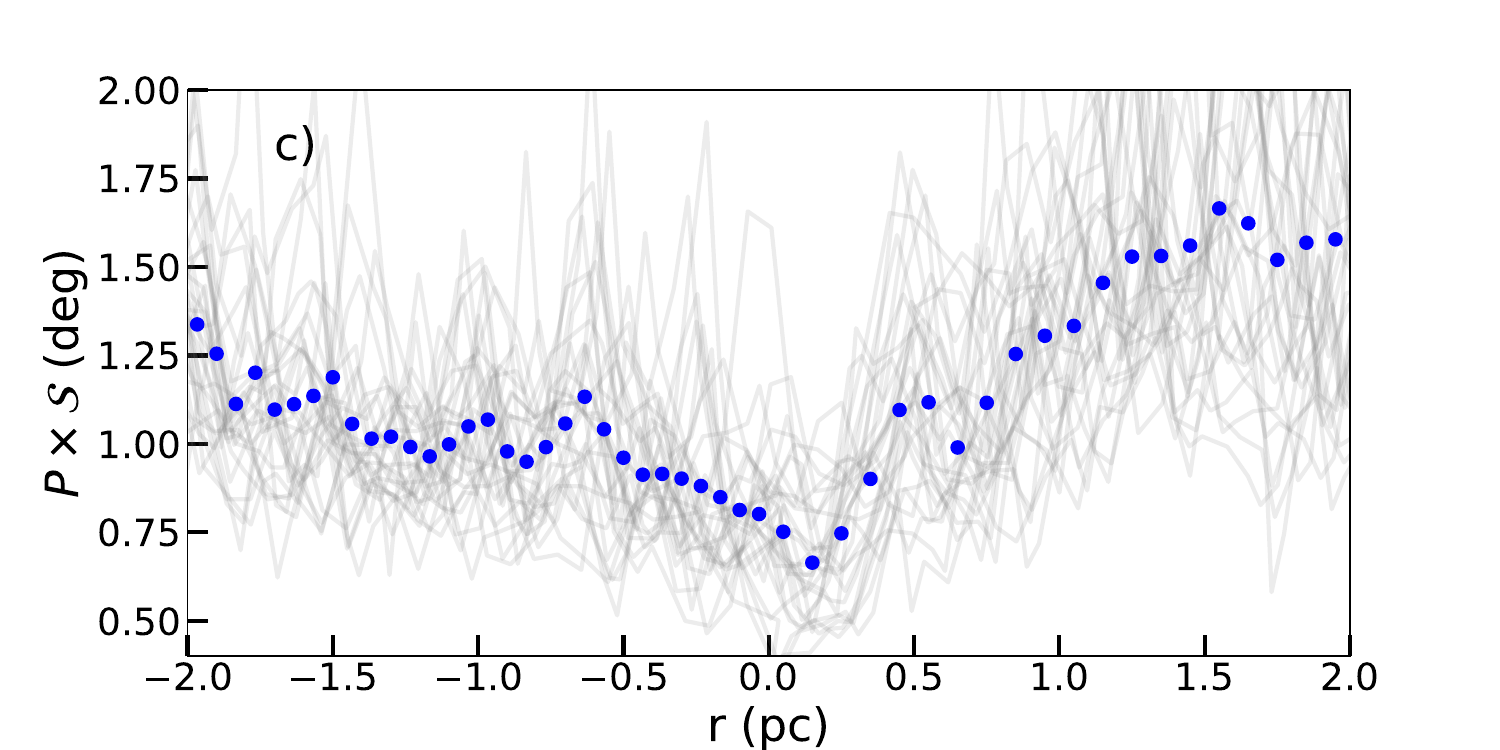}
\includegraphics[trim=0.5cm 0cm 1.5cm 1cm,clip,width=0.47\textwidth]{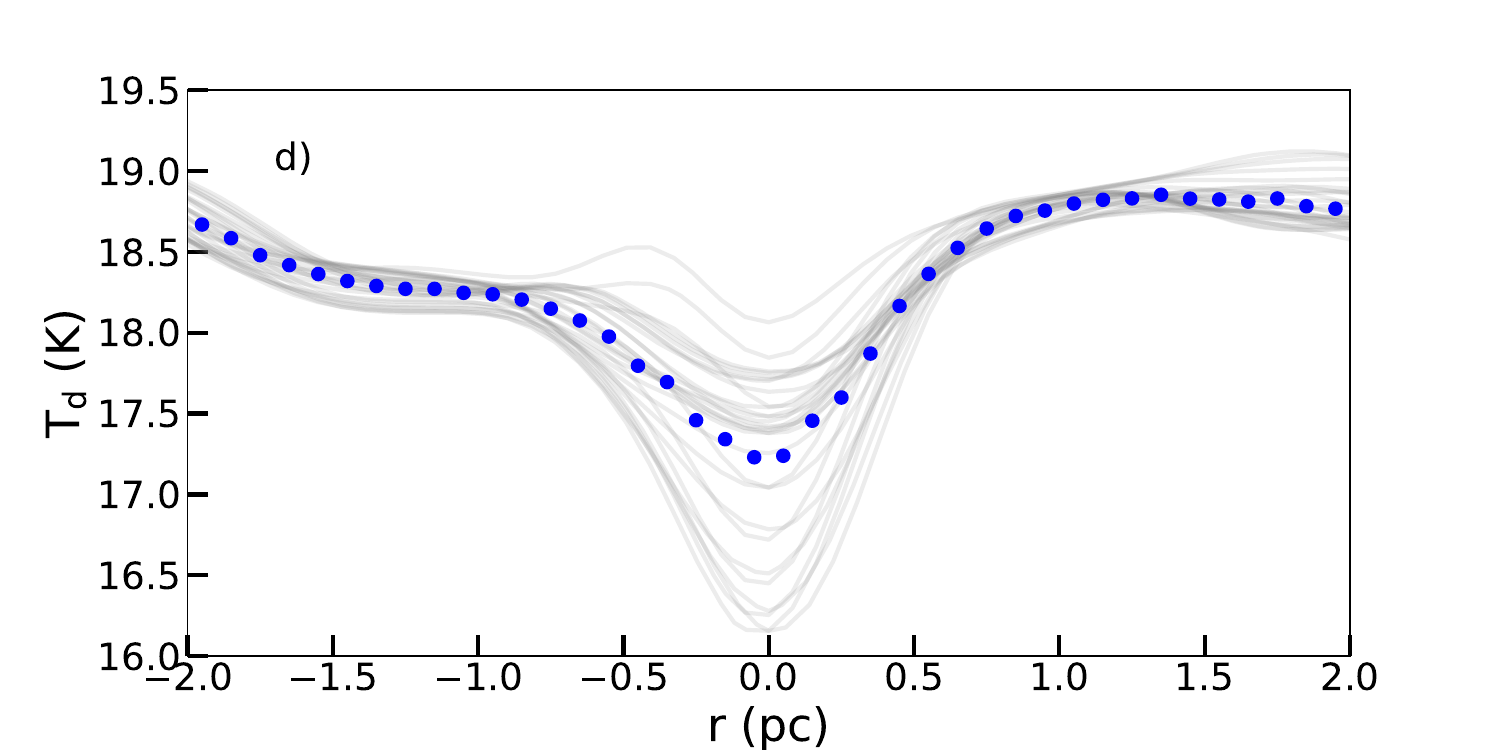}
\caption{Musca. Variations of $P$, $\S$, $P\times \S$, and $T_{\rm d}$ as a function of the radial distance ($r$) constructed using \textit{RadFil} by sampling radial cuts every 0.01 pc along the spine (similar to Figure \ref{fig:musca_radfil}). The radial distance is then defined as the projected distance; the negative distance is to the left, and the positive is to the right. Thin gray lines show the radial cuts, and the mean values are shown as dot makers per bin size of 0.1 pc.}
\label{fig:musca_profiles} 
\end{figure*}

Musca (Figure \ref{fig:musca_maps}) is located at a distance of $170 \pc$ \citep{zucker2021}. The filament is in its early evolutionary stage and lacks active star formation \citep{cox2016filamentary}. One T Tauri star candidate is located at the end of its north \citep{vilas1994dense} (in the galactic coordinates). Musca is known for being an isothermal filament in hydrostatic equilibrium and already fragmented; new stars can be formed in the near future \citep{kainulainen2016musca, bonne2020formation}. Polarization measurements from starlight \citep{Pereyra2004musca} and thermal dust emission (\citealt{Planck_2016_33, cox2016filamentary}) show that local B-fields are generally perpendicular to the filament's spine. Using {\it Herschel} data, \citet{cox2016filamentary} found that B-fields follow the low-column density striations. It suggested that Musca has been accreting interstellar material through these low-column density striations. Recently, B-fields in a small part of the filament obtained by SOFIA/HAWC+ at higher spatial resolution also indicate that the B-field orientation is generally perpendicular to the filament's spine \citep{Kaminsky2023_musca3d}.

We use the polarization map from {\it Planck} at $\sim$$850\;\mum$ (353 GHz). This is the highest frequency and most sensitive channel of {\it Planck} to dust polarization \citep{Planck_2016_33}. The Stokes $I$, $Q$, and $U$ are mapped at the resolution of $5\arcmin$ corresponding to a physical scale of $\sim$$0.24 \pc$. The pixel size of the maps is $112\arcsec$. Musca is one of the high signal-to-noise ratio (S/N) regions of the {\it Planck} full-sky polarization map \citep{Planck_2016_33}. Figure \ref{fig:musca_maps} (left) displays the B-field map of Musca from {\it Planck} with the lengths of line segment proportional to the polarization fraction ($P$). The map is extracted from the {\it Planck} map and then reprojected to the galactic coordinates (using the \textit{reproject} Python package of \textit{astropy}).

Figure \ref{fig:musca_maps} (center and right) show the $\NHt$ and $\Td$ maps obtained from the SED fitting by \citet{Planck_11_mbb}. The density map reveals a nice filamentary structure with several fragments along the filament, and the column density increases from the filament's outer to inner parts. The dust temperature map shows two trends in the filament's spine: high temperature in the northern part $\sim$$21\;\K$ and low temperature in the southern one $<17\;\K$. However, we note that the dust temperature of the filament from {\it Planck} data, which can go up to $\sim$$22\;\K$, is higher than that obtained by {\it Herschel} with a maximum $\Td$ of $\sim$$18\;\K$. Moreover, \cite{bonne2020formation} reported using {\it Herschel} data that there is no high-temperature core as seen with the {\it Planck} temperature map (Figure \ref{fig:musca_maps} right). Be noted that the {\it Planck} dust temperature is obtained based on six channels ranging from 20 to 3000 $\mu$m with a representative spatial resolution of $4\arcmin.9$ at 850 $\mu$m while for {\it Herschel} it is five channels ranging from 70 to 500 $\mu$m with much higher spatial resolution such as $36\arcsec$ at 500 $\mu$m. The differences of the resulting $\Td$ maps may come from the different dust components probed by {\it Planck} and {\it Herschel}. Moreover, the high dust temperatures from {\it Planck} may be due to the confusion between the cold and warm components of the SED fitting (Lars Bonne in private communication). With Musca, we aim to study the non-embedded radiation source regions; for further studies, we only work with the southern part of the filament encompassed by the black rectangle in Figure \ref{fig:musca_maps} (right).

\citet{Planck_2016_33} conducted a detailed study of Musca using polarization data obtained by {\it Planck}. Their findings revealed a depolarization effect within the filament, manifesting as a drop in polarization fraction in the filament's spine and a decrease in the polarization fraction with increasing gas column density. To understand the cause of the depolarization, we will examine the filament's density, temperature, and B-field tangling. Due to its well-defined filamentary structure, we will explore the radial profiles of those quantities and their relations.

\subsubsection{Gas Density Profile}
\label{subsection:muscagasprofiles}

We fit the radial density profile of Musca with a Plummer-like function using a Python package \textit{RadFil} applied for a column density map \citep{zucker2018radfil} (more details can be found in Appendix \ref{appendix:muscaradfil}). Assuming that the filament is in the POS, the Plummer-like function is as follows \citep{arzoumanian2011filament, cox2016filamentary, zucker2018radfil}:
\bea
N({\rm H_2})(r)=\frac{N({\rm H_2})(0)}{[1+(r/R_{\rm flat})^2]^{\frac{q-1}{2}}} \label{eq:plummer}
\ena
where $r$ is the radial distance from the filament's spine; $R_{\rm flat}$ is the radius of the inner, flat region of the density profile; $N({\rm H_2})(0)$ is the peak column density, and $q$ is the power-law density exponent. 

With the best-fit model given by \textit{RadFil}, we find $q = 1.94$ and $R_{\rm flat} = 0.28 \pc$. \citet{cox2016filamentary} fitted a Plummer-like function for Musca using {\it Herschel} data and obtained $q = 2.2$ and $R_{\rm flat} = 0.08 \pc$. Because {\it Planck}'s beam size of $\sim$$5 \arcmin$ ($\sim$$0.24 \pc$) is much larger than that of {\it Herschel} of $\sim$$36\arcsec$ ($\sim$$0.025 \pc$), our fitted $R_{\rm flat}$ is much larger than that of {\it Herschel}.

\subsubsection{Filament Radial Profiles}
\label{subsection:muscaprofiles}

Figure \ref{fig:musca_profiles} shows the physical parameters across the filament obtained by \textit{RadFil}, including polarization fraction (a), polarization angle dispersion function (b), grain alignment efficiency $P\times \S$ (c), and dust temperature (d). 

It is seen from the $P$ profile (a) that $P$ is the lowest in the filament's spine ($\sim$8\%). $P$ increases to 15\% at about $0.7\;\pc$ from the spine and then slightly decreases to 10\% when going to the ISM. This trend is also found by \cite{Planck_2016_33} (see their Figure 10). 

To unravel the effects of B-field tangling on the decrease of $P$ with decreasing $r$ (depolarization), we analyze the polarization angle dispersion function, $\S$, and the product $P \times \S$ shown in Figure \ref{fig:musca_profiles} (b and c). $\S$ is generally low in Musca, and its mean value is smaller than $10^\circ$. \citet{Planck_2016_33} claim that this depolarization is not attributed to the B-field fluctuations because $\S$ is small ($\sim$$10^\circ$), significantly smaller than $52^\circ$ for random fields.

In addition to B-field tangling, the polarization fraction is influenced by environmental factors and alignment efficiency. Removing the effect of the B-field tangling, $P \times \S$ represents the alignment efficiency of grains along the LOS \citep{Planck2020p12}. The $P \times \S$ radial profile reveals that the alignment efficiency is the lowest in the filament spine and increases when moving from the inner to the outer filament.

The temperature profile (d) shows that the temperature and column density (Figure \ref{fig:musca_radfil}) behave in the opposite manner; the temperature is low in the filament's spine, which has the highest density ($\sim$$5 \times 10^{21} \cm^{-2}$). The $\NHt-$$T_{\rm d}$ anti-correlation was also found with {\it Herschel} data by \cite{bonne2020formation}. The temperatures in the spine are below $18\;\K$ and can go down to $16\;\K$. The mean temperatures increase from $17\;\K$ to $20\;\K$ from the spine to the outer areas.

\begin{figure*}[!htb]
\centering
\includegraphics[trim=0.2cm 0.5cm 0.5cm 1cm,clip,width=6.18cm]{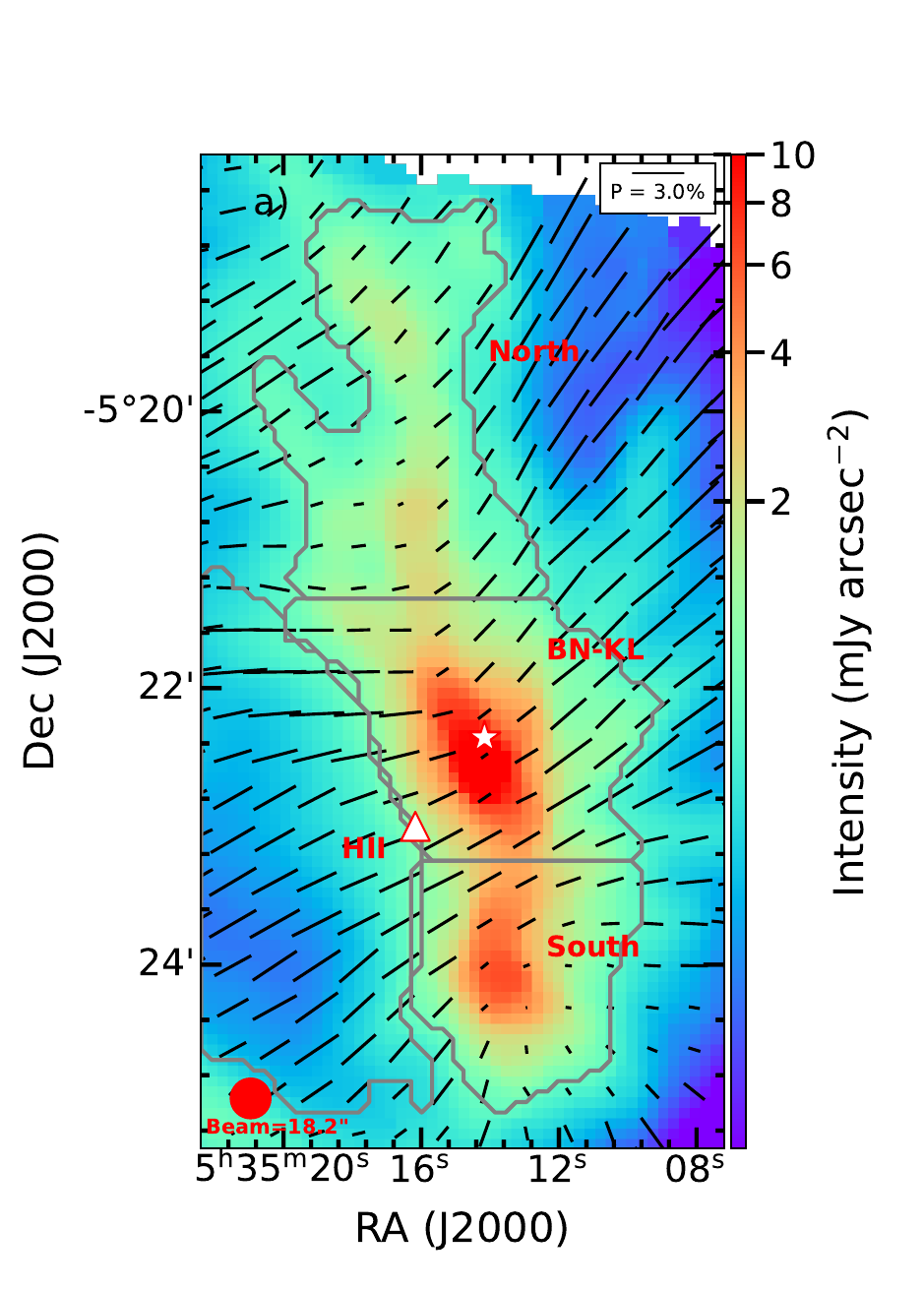}
\includegraphics[trim=0.cm 0.cm 0.cm 0.cm,clip,width=5.7cm]{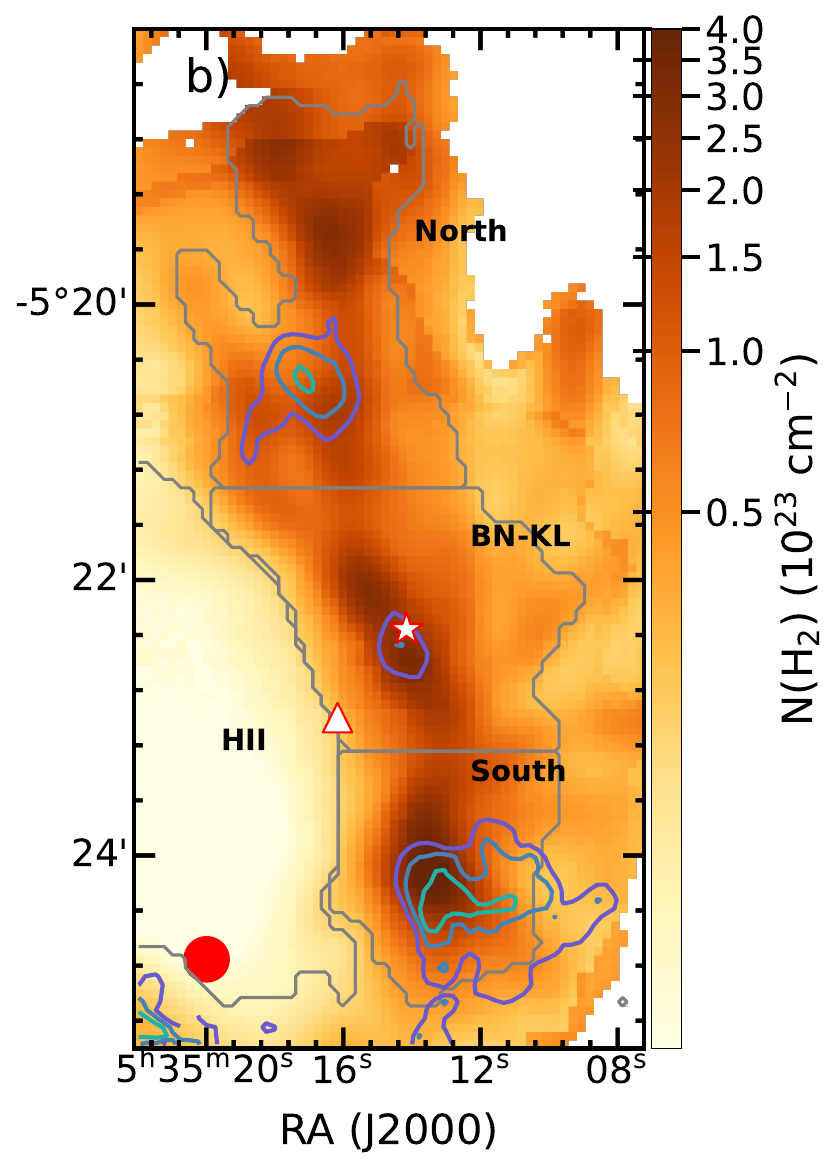}
\includegraphics[trim=0.cm 0.cm 0.cm 0.cm,clip,width=5.72cm]{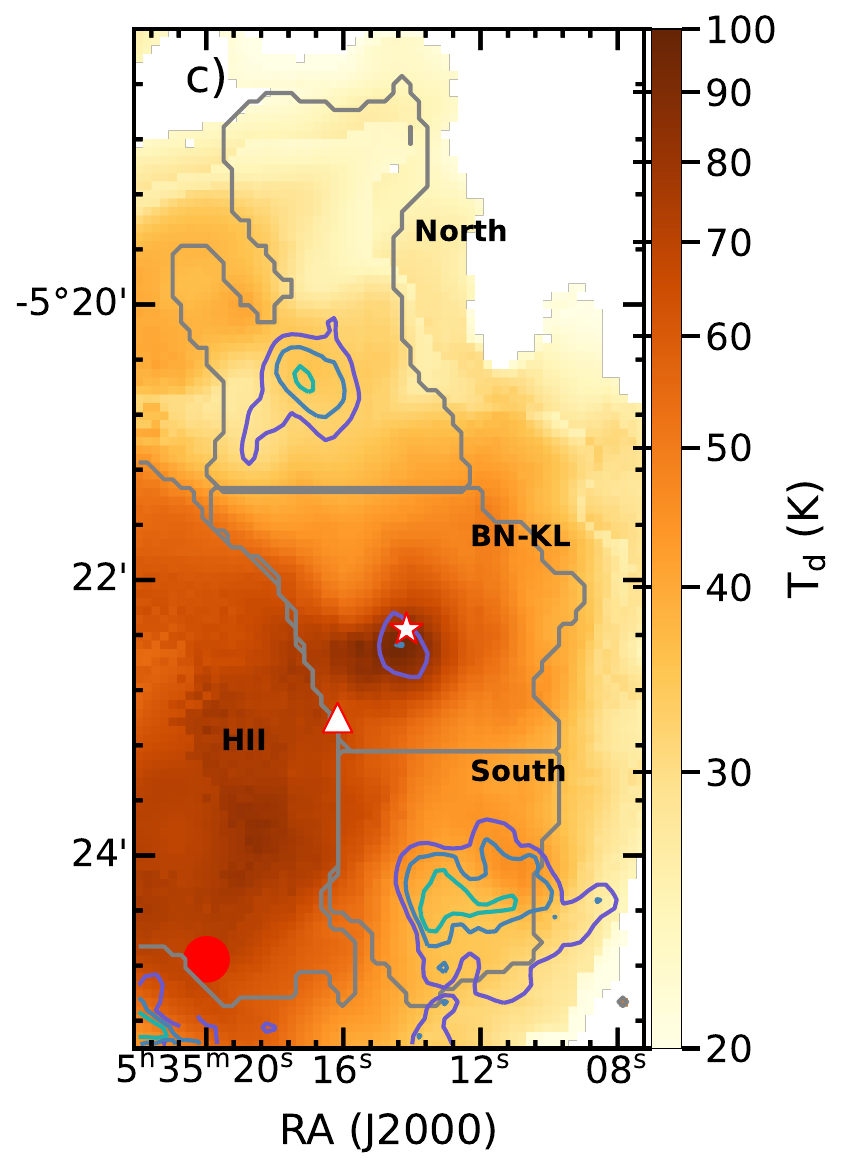}
\caption{OMC-1. a) B-field orientation map observed at $214 \,\mum$ wavelength by SOFIA/HAWC+. The black lines show the orientation of the B-field vector and are overplotted on the intensity map. The vector spacing equals the beam size, shown by a solid red circle in the lower-left corner. Three regions, North, BN-KL, and South, are labeled. b) the column density map, and c) the dust temperature map. The light green, dark blue, and violet contours correspond to $P = 0.6,1.0,1.4 \%$. The star indicates the location of BN/KL, and the triangle shows the location of the Trapezium cluster.} \label{fig:orionmaps} 
\end{figure*}

From those filament profiles, we see that the polarization fraction is lowest in the filament bone in which the temperature (or radiation field) is lowest, density is highest, and alignment efficiency is the lowest. The polarization fraction increases when density decreases, dust temperature increases, and alignment efficiency increases.

\begin{figure*}[!htb]
\centering
\includegraphics[trim=0.cm 0.cm 1.5cm 0.cm,clip,width=0.45\textwidth]{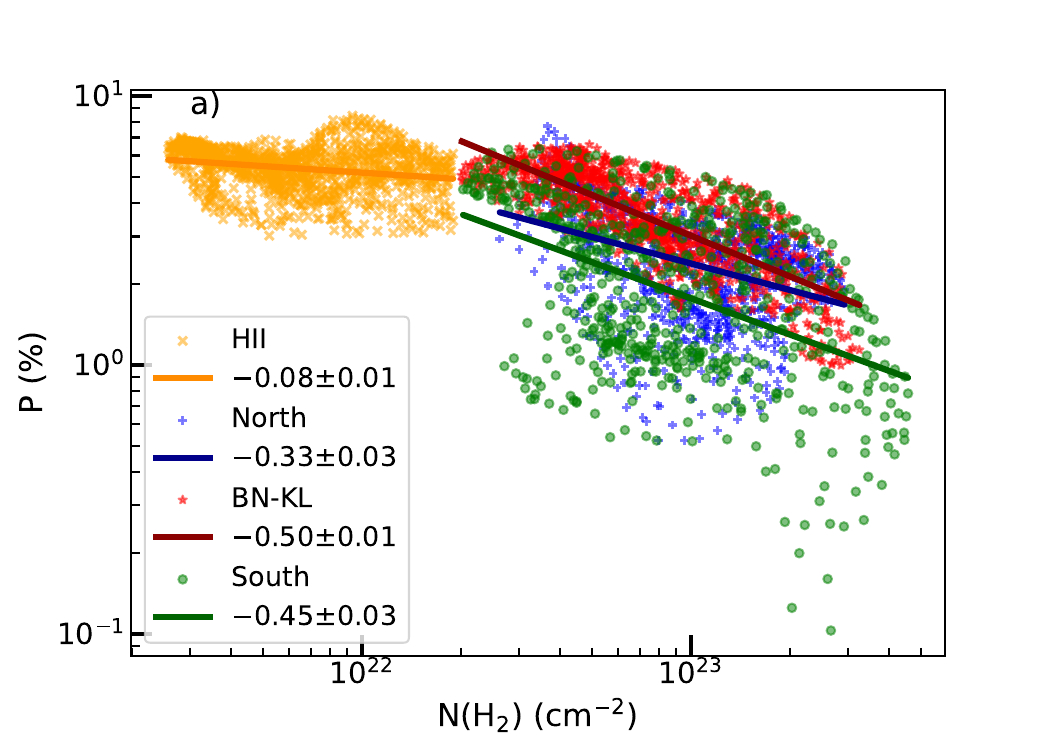}
\includegraphics[trim=0.cm 0.cm 1.5cm 0.cm,clip,width=0.45\textwidth]{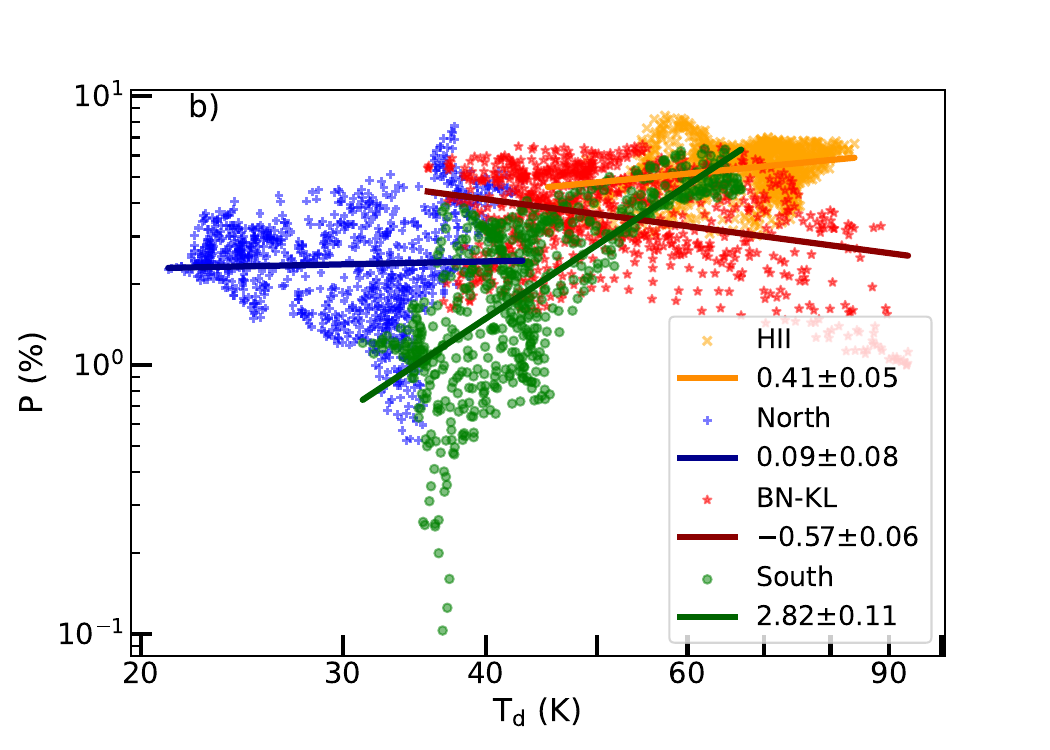}
\includegraphics[trim=0.cm 0.cm 1.5cm 0.cm,clip,width=0.45\textwidth]{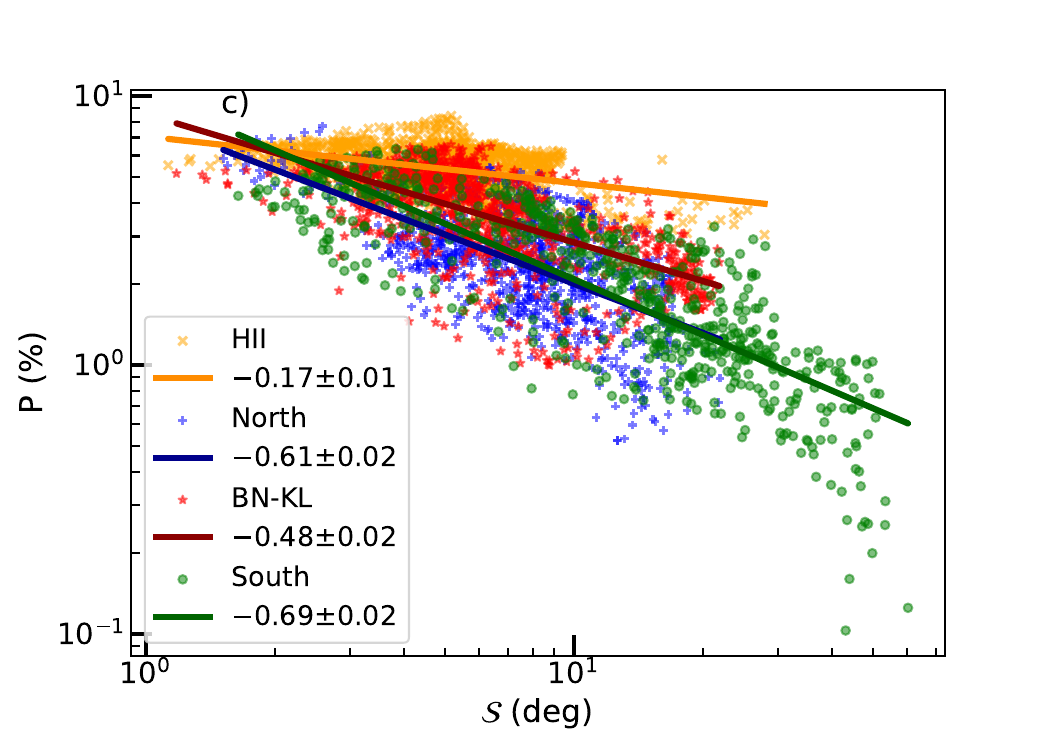}
\caption{OMC-1. Variations of the polarization fraction, $P$, with the column density $\NHt$ (a), dust temperature $\Td$ (b), and angle dispersion function $\S$ (c). The blue, red, green, and orange dots represent the North, BN-KL, South, and the \ion{H}{2} region and the lines are the results of fits to power-law models, respectively.} \label{fig:orionrelation} 
\end{figure*}

\begin{figure*}[!htb]
\centering
\includegraphics[trim=0.cm 0.cm 1.5cm 0.cm,clip,width=0.45\textwidth]{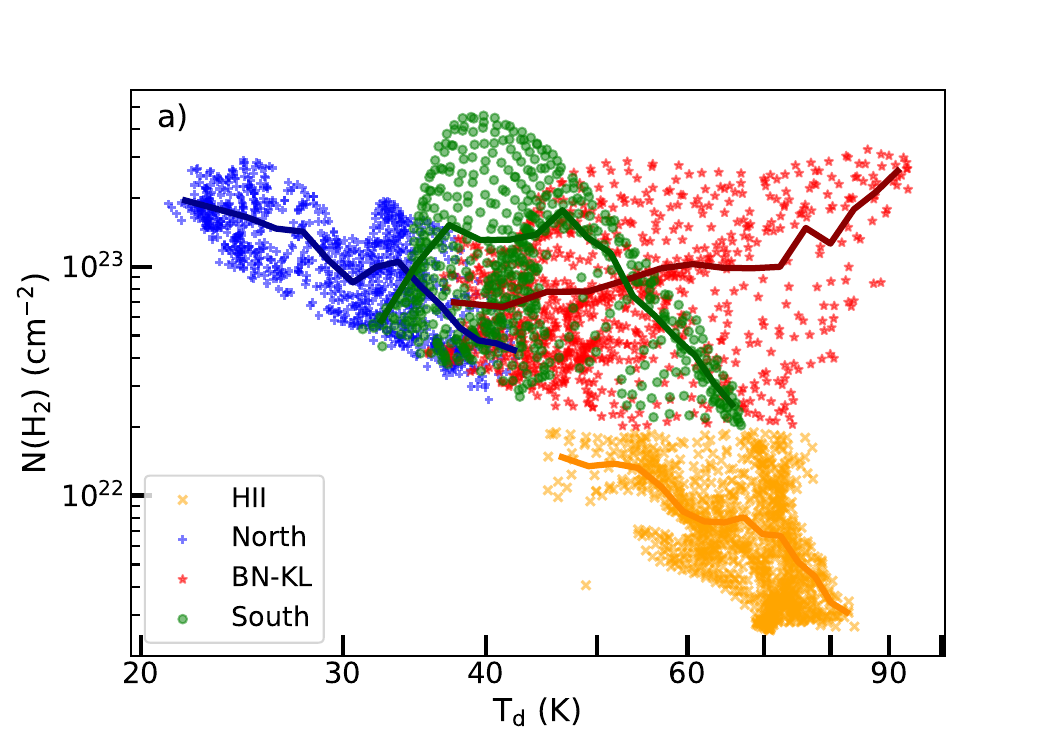}
\includegraphics[trim=0.cm 0.cm 1.5cm 0.cm,clip,width=0.45\textwidth]{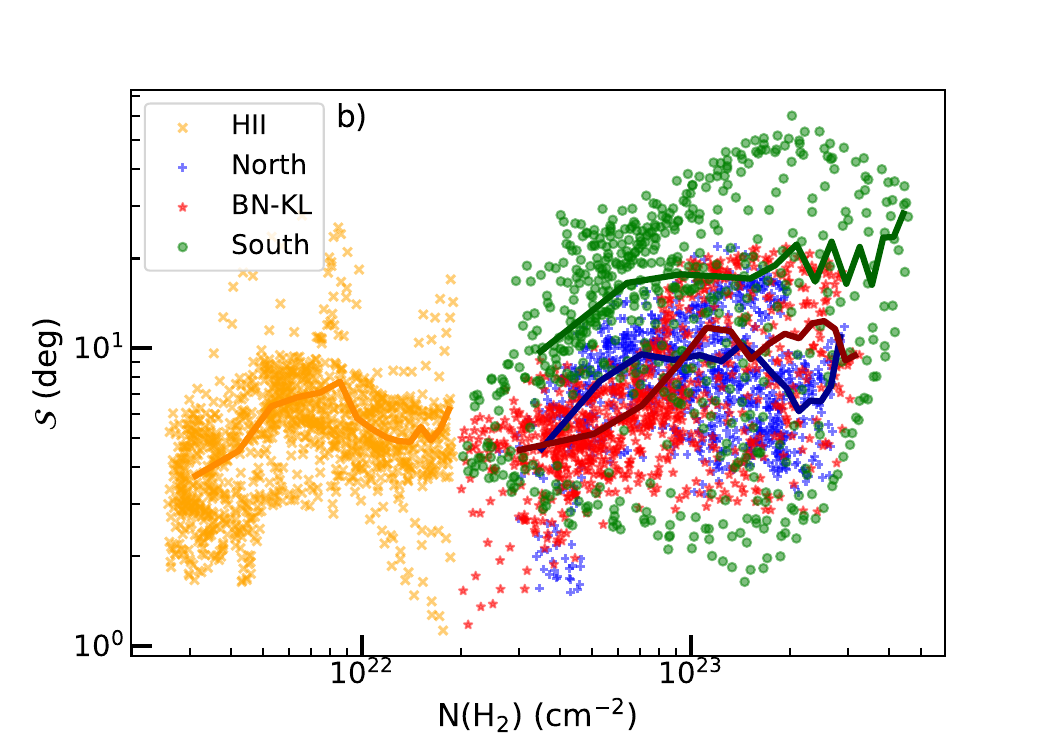}
\includegraphics[trim=0.cm 0.cm 1.5cm 0.cm,clip,width=0.45\textwidth]{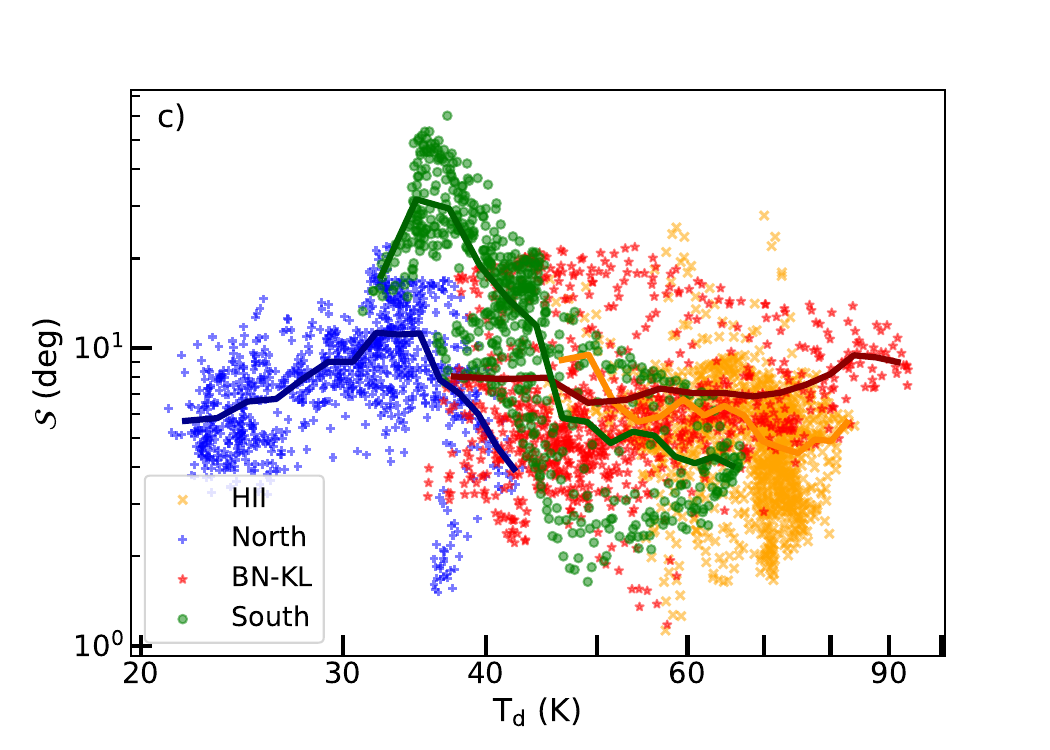}
\includegraphics[trim=0.cm 0.cm 1.5cm 0.cm,clip,width=0.45\textwidth]{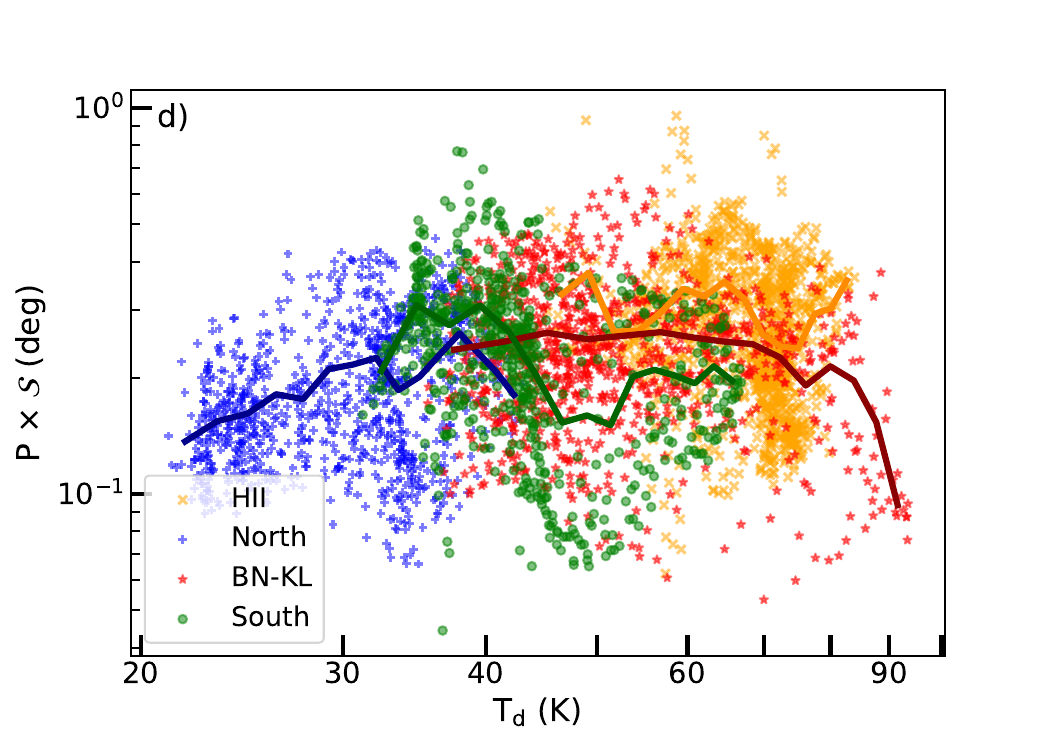}
\caption{OMC-1. The dependence of column density on dust temperature (a), the angle dispersion function $\S$ on the column density $\NHt$ (b) and dust temperature $\Td$ (c), and the dependence of the $P \times \S$ on the dust temperature $\Td$ (d). The blue, red, green, and orange dots represent the data in the North, BN-KL, South, and the \ion{H}{2} regions, respectively, and the solid curves show the running mean values of the data.} \label{fig:orionrelation2} 
\end{figure*}

\subsection{OMC-1}
\label{subsec:obsorion}

Located at a distance of $388 \pm 5 \pc$ \citep{kounkel2017oriondistance}, the Orion Nebula is the nearest high-mass star formation region \citep{odell2008orionnebula}. OMC-1 or Orion A is located behind an \ion{H}{2} region ionized by O-B stars from the Trapezium cluster. OMC-1 consists of two principal clumps, the northern Becklin-Neugebauer-Kleinmann-Low (BN/KL) clump \citep{becklinneugebauern1967, kleinmannlow1967} and the southern Orion S clump \citep{batria1983orions}. BN/KL hosts an extremely explosive molecular outflow with a wide opening angle and multiple ejecta. B-fields of OMC-1 have an hourglass morphology and strengths of the order of mG (e.g., \citealt{pattle2017jcmt,chuss2019hawc+, guerra2021orion}).

We use thermal dust polarization data observed toward the Orion BN/KL by SOFIA/HAWC+ at $214\,\mum$ and beam size of $18\arcsec.2$, which was firstly reported by \citet{chuss2019hawc+}. Analyzing this data set, \citet{chuss2019hawc+} and later \citet{guerra2021orion} mainly focused on the studies of B-fields. \citet{chuss2019hawc+} found the depolarization effect. Evidence of RAT-D is found by \citet{tram2021bnkl} in BL-KL, and by \cite{LeGouellec2023} in the Orion Bar. For the present work, we only use the data having signal-to-noise ratios ($S/N$s) satisfying: $S/N(I)>250$ and $S/N(P)>3$.

Figure \ref{fig:orionmaps} displays the maps of B-fields (a), column density (b), and dust temperature (c) observed toward OMC-1 by SOFIA/HAWC+ at 214 $\mum$. We only consider the OMC-1 filament and \ion{H}{2} region encompassed by the gray contours. The \ion{H}{2} region has low column density of $\NHt < 2 \times 10^{22} \cm^{-2}$, and high dust temperature $T_{\rm d}> 45 \K$. While the filament has high column density $\NHt >2 \times 10^{22} \cm^{-2}$ and strong emission $I>15$ $\rm mJy/arcsec^{-2}$; for a detailed study, we divided the OMC-1 filament into three sub-regions: North, BN-KL, and South.

The $\NHt$ map (Figure \ref{fig:orionmaps} b) shows that the column density is highest in the filament's spine and divided into three clumps residing in the North, BN-KL, and South regions. BN-KL exhibits the highest temperatures, reaching 100 K (Figure \ref{fig:orionmaps} c). The temperatures of the eastern part of OMC-1 and the \ion{H}{2} region are higher than that of the western one. This is due to the effect of ionization from the Trapezium cluster. The \ion{H}{2} region has the lowest densities compared to other regions. 

The lengths of the line segments in Figure \ref{fig:orionmaps} (a) are proportional to the polarization fraction, revealing high polarization fractions in the \ion{H}{2} region and particularly low polarization fractions in certain parts of the filament. The color contours in Figure \ref{fig:orionmaps} (b) and (c) highlight three polarization holes in the central part of each sub-region, with polarization fractions smaller than 1\%. In Figure \ref{fig:orionmaps} (b) and (c), the polarization holes in the North and South regions show lower dust temperatures compared to the polarization hole in BN-KL, despite all regions having high column density. This suggests that the depolarization in the low and high dust temperatures may be caused by different mechanisms.

To explicitly see how the polarization fraction, $P$, changes with the column density, $\NHt$, dust temperature, $\Td$, and B-field tangling, $\S$, we show the variations of $P$ with $\NHt$, $\Td$, and $\S$ in Figure \ref{fig:orionrelation}. 
The $P-\NHt$ relation (Figure \ref{fig:orionrelation} a) shows an anti-correlation of the polarization fraction and the gas column density (i.e., polarization hole). We fit the data with a power-law model and obtain the power indices $\alpha = - 0.33 \pm 0.03$ for North, $\alpha = - 0.50 \pm 0.01$ for BN-KL, $\alpha = - 0.45 \pm 0.03$ for South, and the smallest slope value $\alpha = - 0.08 \pm 0.01$ for the \ion{H}{2} region. In the OMC-1 filament, the $P-\Td$ relation reveals that the polarization fraction increases at low $\Td$ and dramatically decreases as the dust temperature increases to $\Td > 70 \K$. In detail, the polarization fraction remarkably increases with increasing $\Td$ with a power index $\alpha = 2.82 \pm 0.11$ in the South, while in the North with the low temperature ($\Td < 40 \K$), the polarization fraction slightly increases with $\alpha = 0.09 \pm 0.08$. In BN-KL, having massive protostars and a high dust temperature region, the polarization fraction only decreases as the dust temperature increases with $\alpha = - 0.57 \pm 0.06$. \citet{tram2021bnkl} studied this relation for BN-KL using data collected by JCMT/POL-2 and SOFIA/HAWC+ and found this increase-decrease feature. Although dust temperature can go up to 80 K, the polarization fraction gradually increases with increasing $\Td$ in the \ion{H}{2} region with $\alpha = 0.41\pm 0.05$.

We also fit a power law function to the $P-\S$ relation and find the anti-correlation of $P$ and $\S$ with $\alpha = -0.61 \pm 0.02, -0.48 \pm 0.02$, $-0.69 \pm 0.02$, and $-0.17\pm0.01$ for North, BN-KL, South, and \ion{H}{2} region, respectively. The \ion{H}{2} region seems to have weak B-field fluctuations as shown by the small angle dispersion function of $\S<10^\circ$. However, the South region appears to have stronger B-field fluctuations due to the high angle dispersion function of $\S>10^\circ$.

From the $\NHt$$-T_{\rm d}$ relation shown in Figure \ref{fig:orionrelation2} (a), we see that $\NHt$$-T_{\rm d}$ is anti-correlated in the North, South, and \ion{H}{2} regions, while, it is correlated in the BN-KL region with the presence of massive protostars. 
To show the effect of B-field tangling on the depolarization, we plot the relations of $\S-\NHt$ and $\S-\Td$ in Figure \ref{fig:orionrelation2} (b) and (c). The $\S-\NHt$ relation shows the correlation of the polarization angle dispersion function $\S$ and the column density in general. This clearly shows the increasing importance of B-field tangling in dense regions. The $\S - \Td$ relation shows that the polarization angle dispersion function $\S$ has anti-correlation with the dust temperature in the South and \ion{H}{2} regions, but a correlation in the North and BN-KL regions. The relationship between $P \times \S$ and $\Td$ (Figure \ref{fig:orionrelation2} d) shows that the average grain alignment efficiency firstly slightly increases with dust temperature for $\Td < 35 \K$, remains constant in the range of 35-70 K, and then decreases for $\Td > 70 \K$.


\section{Numerical Modeling and Results}
\label{sec:model}

This section presents the numerical modeling method of polarized thermal dust emission from aligned grains using the RAT theory for Musca and OMC-1 to test the RAT paradigm. We first use the DustPOL-py code\footnote{\url{https://github.com/lengoctram/DustPOL-py}}, which was first presented in \cite{lee_2020} and improved by \cite{tram2021ophiuchi} to obtain the dust polarization for the ideal model in which the B-fields are assumed to lie in the POS. Then, we incorporate the depolarization effects caused by the B-field inclination angle and their fluctuations into our ideal model to obtain a realistic polarisation model and confront it with observational data.

Below, we briefly describe the main features of DustPOL-py  and its key input parameters, including the physical parameters of the gas and dust, the local radiation field, and the alignment of dust grains by RATs. We then present the main modeling results and compare them to the data. Our modeling approach is physical modeling based on the alignment physics of RATs. There exist other polarization modeling approaches, such as inverse modeling \citep{kim1995size,Draine2006,Hoang2013} and parametric modeling \citep{Guillet2018, Hensley.2023}. However, the latter two modeling methods are not based on underlying physics and require multi-wavelength data to derive model parameters; therefore, they are not convenient for pixel-by-pixel modeling with single-wavelength data, which we do in the present study.

\subsection{Methods and Assumptions}

\subsubsection{Gas Properties and Radiation Fields}

The gas density ($n_{\rm H} =  2n_{\rm H_2}$) and temperature ($T_{\rm gas}$) are the two important physical parameters in the RAT alignment theory because they determine the randomization and alignment of dust grains (see e.g., \citealt{hoang2021polhole}). For our modeling, these parameters are inferred from observational data and used as the input parameters for DustPOL-py .

For radiation fields, the radiation strength, $U$ (effectively equivalent to the dust temperature $\Td$) and the anisotropy degree of the radiation, $\gamma$, are the key parameters of the RAT theory because they determine the alignment degree of dust grains \citep{hoang2021polhole}. The anisotropy degree varies in a range from 1 to 0; $\gamma=1$ represents a unidirectional radiation field from a nearby star, while $\gamma=0.1$ for the diffuse ISM \citep{draine1996}. The mean wavelength, $\bar{\lambda}$, depends on specific regions.


\subsubsection{Grain Size Distribution}
We adopt a power-law grain size distribution, $dn/da \propto a^{-\beta}$, where $\beta$ is in the range of $3.5-4.5$. The grain size distribution in the ISM follows the MRN distribution with a power-law index $\beta=3.5$ \citep{mathis1977size}. The smallest grain size, $a_{\rm min}$, is 10$\Angstrom$, while the initial maximum grain size, $a_{\rm max}$, can be varied in the range of $0.2\;\mum$ to several microns.

When considering the RAT-D effect, the value of $a_{\rm max}$ varies with the local conditions of the gas densities, radiation fields, and tensile strengths, as given by Equation \ref{eq:adisr}. Their tensile strengths determine the internal structure of grains, and the disruption temperature depends on this property. For composite grains, the tensile strengths vary around $S_{\rm max} = 10^{7} \erg \cm^{-3}$ \citep{Hoang2019ApJ}.

\subsubsection{Grain Alignment Function by RATs}

In the RAT-A theory, grains are first spun up to suprathermal rotation and then driven to be aligned with the ambient B-fields by RATs \citep{lazarianhoang2007,hoang2008radiative}. For grains with iron inclusions, which are superparamagnetic material, the joint effect of enhanced magnetic relaxation and RATs could make grains achieve perfect alignment with $\Bv$, a mechanism known as magnetically enhanced RAT (MRAT) \citep{hoang2016unified}. Therefore, grains are only efficiently aligned when they can rotate suprathermally. The minimum size for grain alignment, $a_{\rm align}$ (also called critical alignment size), is determined by the suprathermal rotation criterion of $\omega_{\rm RAT}(a_{\rm align})=3\omega_{\rm T}$ \citep{hoang2008radiative}, where $\omega_{\rm RAT}(a_{\rm align})$ is the angular velocity of grains spun-up by RATs and $\omega_{\rm T}$ is the grain's thermal angular velocity. From Equation (\ref{eq:angvelo_rat}), one can obtain:
\bea
a_{\rm align} &\simeq& 0.024\hat{\rho}^{-5/32}\gamma^{-5/16}U^{-5/16}\left(\frac{10^3 \cm^{-3}}{n_{\rm H}}\right)^{-5/16}\nonumber\\
&&\times \left(\frac{\bar{\lambda}}{0.5\,\mum}\right)^{17/32}\left(\frac{20\;\K}{T_{\rm gas}}\right)^{-5/16}\left(\frac{1}{1+F_{\rm IR}}\right)^{-5/16} \,\mum,\label{eq:a_align}
\ena
where $U$ is the radiation strength of the local radiation field \citep{hoang2021polhole}.

The degree of grain alignment as a function of the grain size can be described by:
\begin{eqnarray}
    f_{\rm align}= f_{\rm max}\left[1-\exp\left(-a/2a_{\rm align} \right)^{-3}\right],
\end{eqnarray}
where $a_{\rm align}$ is given by Equation \ref{eq:a_align}, $f_{\rm max}$ describes the maximum degree of grain alignment, and the negligible alignment of small grains of $a\ll a_{\rm align}$ is disregarded \citep{hoang2016unified}. 

According to the MRAT mechanism, the exact value of $f_{\rm max}$ depends on the ratio of the magnetic relaxation to the gas damping rate, which is a function of the grain size, magnetic susceptibility, and local gas density, as shown in \cite{hoang2016unified}. They also numerically demonstrated that grains with embedded iron clusters (i.e., superparamagnetic grains) can achieve perfect alignment of $f_{\rm max}=1$. Moreover, inverse modeling of thermal dust polarization observed by Planck in \cite{Hensley.2023} found that perfect alignment of large grains is required to reproduce the observational data. Therefore, in this paper, we assume $f_{\rm max =1}$, which is expected from the MRAT theory for composite grains containing embedded iron inclusions.

\subsubsection{Ideal Model of Thermal Dust Polarization with DustPOL-py }
\label{subsubsec:polfrac}

Dust grains are heated by starlight and subsequently re-emit radiation in infrared wavelengths. In the assumption of a dust environment containing both carbonaceous and silicate grains, the total emission intensity is calculated as: 
\bea
\frac{I_{\rm em}(\lambda)}{N_{\rm H}} = && \sum_{j = sil,cal} \int_{a_{\rm min}}^{a_{\rm max}} Q_{\rm abs}\pi a^2\nonumber\\
&&\times \int dT B_{\lambda}(T_{\rm d})\frac{dP}{dT}\frac{1}{n_{\rm H}}\frac{dn_j}{da}da,\label{eq:Iem}
\ena
where $Q_{\rm abs}$ is the absorption efficiency, $dP/dT$ is the temperature distribution function, and $B_{\lambda}(T_{\rm d})$ is the Planck function. Here, we disregard the minor effect of grain alignment on total emission intensity (cf. \citealt{HoangTruong23}).

If carbonaceous and silicate grains exist in two separate populations, only silicate grains can be aligned with B-fields due to their paramagnetic nature, while carbonaceous grains (e.g., pure graphite) cannot be efficiently aligned because of their diamagnetic nature \citet{hoang2016unified}.\footnote{Although ideal carbon grains being diamagnetic cannot be aligned with B-fields, hydrogenated carbon grains can be aligned with the B-fields because those grains contain unpaired electrons and become paramagnetic material \citep{Hoang.2023qr}.} However, for a composite dust model (a mixture of silicate and carbon in a dust grain), the composite dust can be aligned with B-fields. Note that observational constraints favor the composite model for interstellar dust, such as Astrodust \citep{Draine2021ApJ} and THEMIS model \citep{Ysard.2024}. Hence, for this study, we assume the composite dust model, so that carbonaceous and silicate grains have the same temperature. 

We consider first the ideal model in which B-fields lie in the POS and the impact of B-field fluctuations is disregarded. The polarized emission intensity from aligned grains is then given by:

\bea
\frac{I_{\rm pol}(\lambda)}{N_{\rm H}} = &&\sum_{j = sil,cal} \int_{a_{\rm min}}^{a_{\rm max}} f_{\rm align}(a)Q_{\rm pol}\pi a^2\nonumber\\
&&\times \int dTB_{\lambda}(T_{\rm d})\frac{dP}{dT}\frac{1}{n_{\rm H}}\frac{dn_j}{da}da,\label{eq:Ipol}
\ena
where $Q_{\rm pol}$ is the polarization efficiency \citep{lee_2020}. 

The polarization degree of thermal dust emission is given by
\bea
P_{\rm mod}^{\rm ideal}(\%) = 100\% \times \frac{I_{\rm pol}}{I_{\rm em}}, \label{eq:P}
\ena
where the subscript ``mod" is for the values calculated from the model and superscript ``ideal" for the ideal polarization degree achievable by grain alignment, assuming the optimal case of B-fields in the POS with the absence of the B-field tangling.

\subsubsection{Realistic Polarization Model: Accounting for B-field Tangling}
For DustPOL-py, B-fields are assumed to lie in the POS, so that the degree of thermal dust polarization could achieve the maximum value, as given by Equation \ref{eq:P}. In a realistic situation, the inclination angle of B-fields and LOS, denoted by $\psi$, contributes to the polarization degree due to the projection effect. Moreover, the fluctuations of B-fields along the LOS cause the depolarization of thermal dust emission, which is described by an additional depolarization parameter $F_{\rm turb}$. $F_{\rm turb}$ is a function of the angle between the local B-field and the mean B-field and is anti-correlated with the polarization angle dispersion function $\S$ \citep{HoangTruong23}. Therefore, the realistic dust polarization degree becomes
\bea
P_{\rm mod}=P_{\rm mod}^{\rm ideal} \sin^{2}\psi F_{\rm turb}.\label{eq:Pmod}
\ena

Numerical simulations in \cite{HoangTruong23} using MHD simulations and polarized radiative transfer in POLARIS found that the polarization model given by Equation \ref{eq:Pmod} is in excellent agreement with synthetic polarization. They also found that $F_{\rm turb}$ decreases with $\S$ as $F_{\rm turb}\propto \S^{-\eta_{1}}$, where $\eta_{1}$ depends on the inclination angle. Numerical simulations in \cite{PlanckXX.2015} showed that there is a correlation between the average of the B-field inclination angle and $\S$, which can be described as $\langle \sin^{2}\psi\rangle \propto \S^{-\eta_{2}}$ (see e.g., \citealt{Hensley2019}). Furthermore, the B-field tangling within the beam size (i.e., B-field fluctuations in the POS) can cause an additional depolarization described by the factor $F_{\rm beam}$ \citep{HoangTruong23}. 
Accounting for all these depolarization effects, the net polarization degree of thermal dust emission can be described by
\bea
P_{\rm mod}=  \Phi P_{\rm mod}^{\rm ideal}\left(\frac{\S}{1^\circ}\right)^{-\eta} ,\label{eq:Pmod2}
\ena
where $\Phi$ is a coefficient that describes the depolarization by the B-field's inclination angle, and $\eta>0$ is the power index describing the depolarization effect caused by B-field fluctuations along the LOS and in the POS, which is expected to vary with the local conditions. In practice, we get the value of $\eta$ from the slope of the $P-\S$ relation. The value $\Phi$ is obtained by fitting the polarization fraction from the model to the maximum polarization observed in each filament. The best-fit values of $\Phi$ are shown in Table \ref{tab:Parameters}.

\subsection{Results}

\begin{table*}[!htb]
\caption{Physical parameters for model}\label{tab:Parameters}
\begin{center}
\begin{tabular}{|c | c |c| c|}
\hline
\hline
  & Parameter & Musca & OMC-1 \\ \hline
 Anisotropic degree & $\gamma$ & 0.3 & 1 \\ \hline
 Mean wavelength & $\bar\lambda \;[\mum]$ & 1.2 & 0.3 \\ \hline
 Gas temperature & $T_{\rm gas}$ [K] & $T_{\rm dust}$ & $T_{\rm dust}$ (filament), 5000\K (\ion{H}{2} region)\\ \hline
 Grain composition & - & Sil+Car & Sil+Car \\ \hline
 Dust mass density & $\rho \; [\g \cm^{-3}]$ & 3 & 3 \\ \hline
 Minimum grain size & $a_{\rm min}\; [\mum]$ & $10^{-3}$ & $10^{-3}$ \\ \hline 
 Maximum grain size & $a_{\rm max}\; [\mum]$ & $0.25-0.5$ & $0.2-2$ \\ \hline
 Power-index & $\beta$ & $-3.5$ & $-3.5$ \\ \hline
 Maximum tensile strength & $S_{\rm max}$ [erg cm$^{-3}$] & $10^7$ & $10^7-10^8$ \\ \hline
 B-field's inclination effect & $\Phi$ & $0.53$ & $0.28$ \\ \hline
\hline
\end{tabular}
\end{center}
\end{table*}

Table \ref{tab:Parameters} summarizes the model parameters for Musca and OMC-1, including the gas number density, $\nH$, gas temperature, $T_{\rm gas}$, ($a_{\rm min}$, $a_{\rm max}$), shape (grain aspect ratio = minor axis/major axis taken to be equal to 1/3), size distribution power index, $\beta$, tensile strength, $S_{\rm max}$, radiation field strength, $U$, the mean wavelength, $\bar{\lambda}$, and the anisotropy degree, $\gamma$.

\subsubsection{Musca}
In Musca, where a filamentary structure is evident and the physical conditions do not vary significantly along the filament, we will model the variation of polarization faction $P$ across the filament, as the function of the radial distance, $r$, from the filament's spine to the outer regions.

\label{subsec:muscamodel}
\begin{figure}[!htb]
\centering
\includegraphics[trim=0.3cm 0cm 1.5cm 0cm,clip,width=8.5cm]{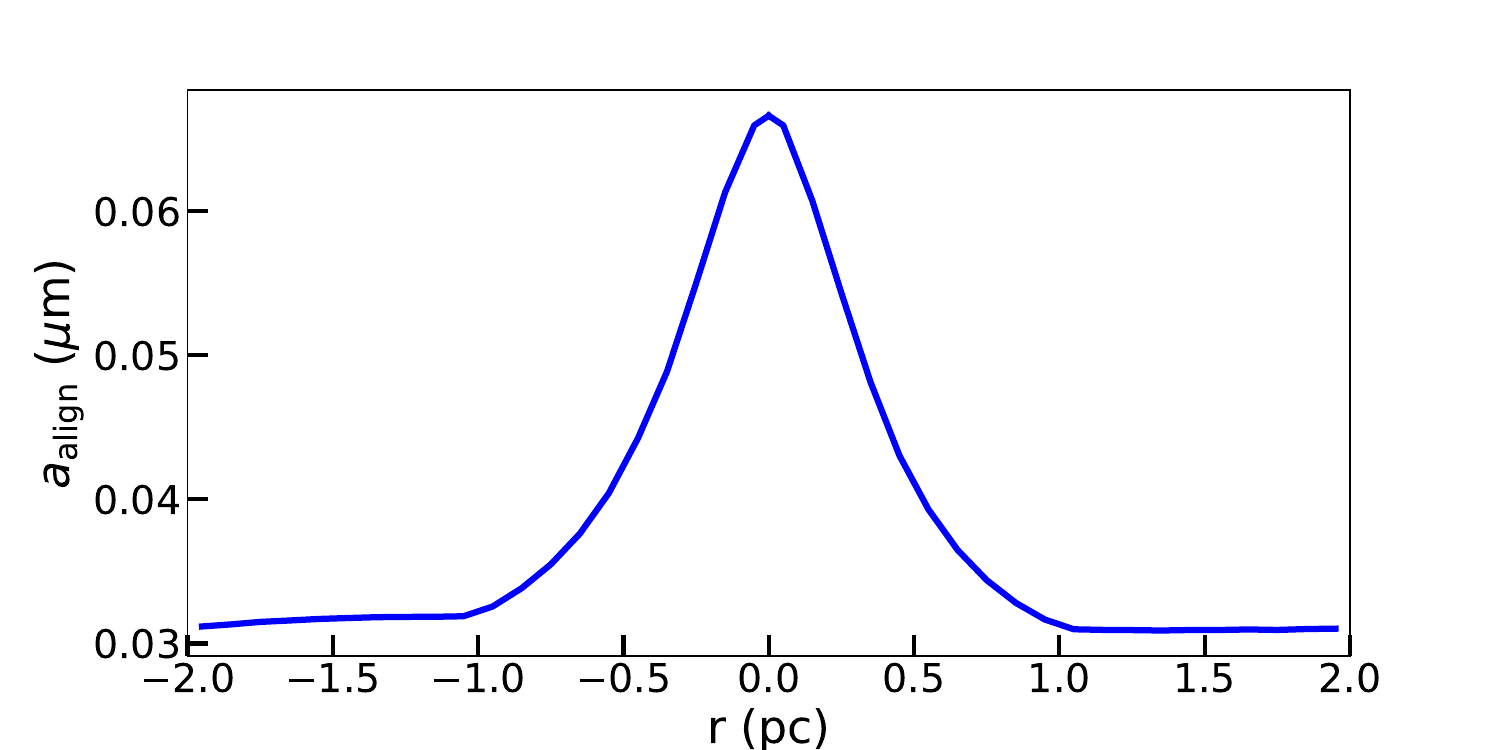}
\includegraphics[trim=0.3cm 0cm 1.5cm 1cm,clip,width=8.5cm]{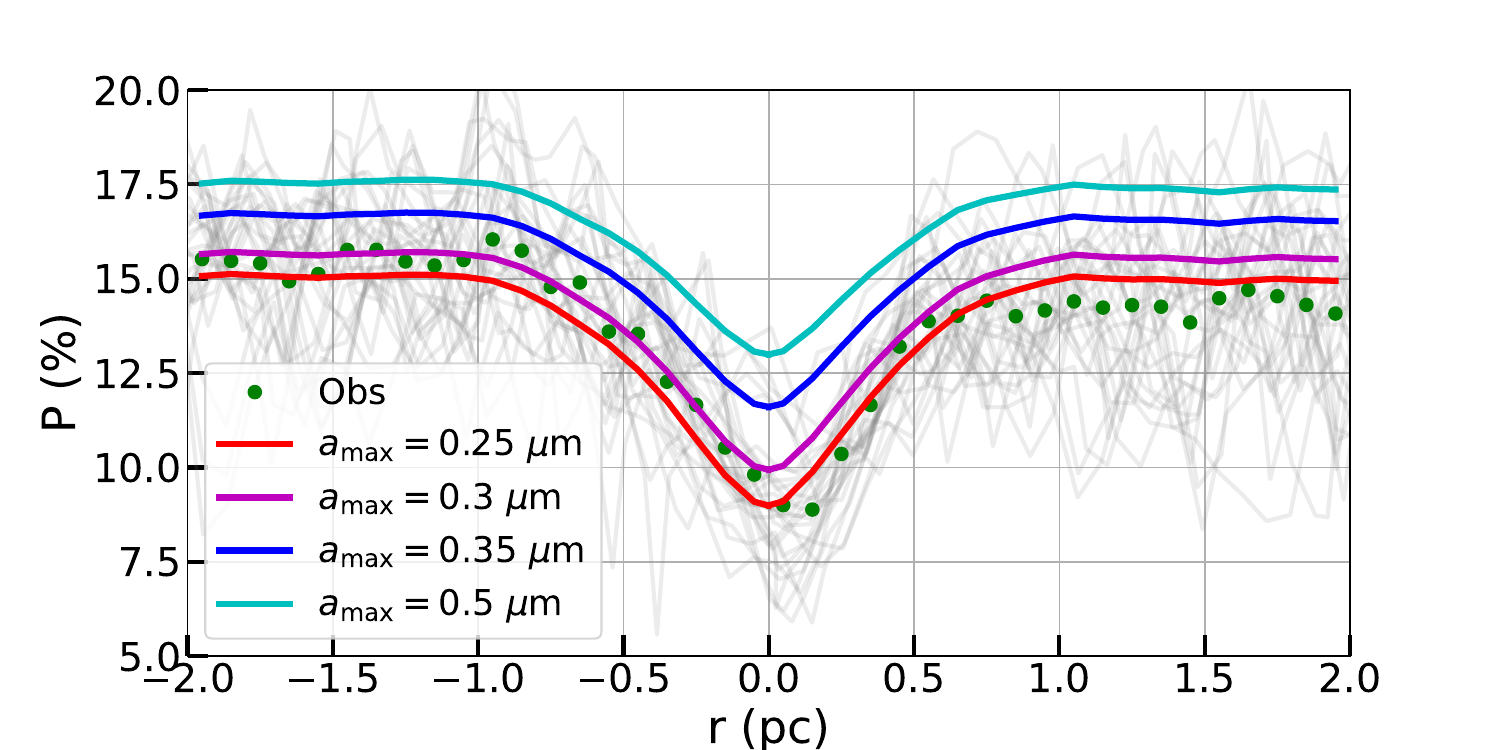}
\caption{Musca. Upper: Minimum alignment size by RATs calculated from our model for Musca. Lower: Comparison of the polarization fraction from our model (thick solid lines) with observation data. The gray curves show $P$ vs. $r$ for radial cuts along the filament's spine from observational data (see Section \ref{subsection:muscaprofiles}). Green dots represent the mean values of the observational $P$ at a certain value of $r$. The color curves show polarization fractions from our model with different $a_{\rm max}$. The rapid increase of alignment size toward the filament center causes the decrease of the model polarization fraction and successfully reproduces the observed polarization hole.} \label{fig:muscamodel} 
\end{figure}

To model the dust polarization as a function of the radial distance, $r$, we first need to know how the local gas volume density and dust temperature (radiation field) vary with $r$. The dust temperature profile is taken to be the mean temperature obtained from {\it Planck} (Figure \ref{fig:musca_profiles} d).

For the regions of $|r|<1$ pc within the filament, the gas number density varies with the radial distance $r$ as the Plummer-like function (e.g., \citealt{arzoumanian2011filament}),
\bea
n_{{\rm H}_2}(r) = \frac{n_{{\rm H}_2}(0)}{[1+(r/R_{\rm flat})^2]^{p/2}},\label{eq:nH_Musca}
\ena
where $n_{\rm H_2}(0)$ is the number density at the filament center $r = 0$, given by
\bea
n_{\rm H_2}(0) = \frac{N({{\rm H}_2})(0)}{A_{\rm p}R_{\rm flat}}, 
\ena
where 
\bea
A_{\rm p} = \frac{1}{\cos i} \int_{-\infty}^\infty \frac{du}{(1+u^2)^{p/2}},
\ena
with $i$ is the inclination angle of the filament relative to the POS, and we assumed $i = 0$. Following Plummer-like fitting as described in Section \ref{subsection:muscaprofiles}, we obtain $n_{\rm H_2}(0) \simeq 1300 \cm^{-3}$. 

For the outer regions of $|r|>1 \pc$, we assume a constant gas density of $n_{\rm H_2} = 100 \cm^{-3}$ and constant dust temperature of $T_{\rm d}=18.3\;\K$ (Figure \ref{fig:musca_profiles} d).

Figure \ref{fig:muscamodel} (upper panel) shows the variation of the minimum alignment size by RATs ($a_{\rm align}$) computed for Musca. The alignment size increases rapidly toward the filament spine due to the increase in the gas volume density and reduction of the radiation field (dust temperature). For a fixed maximum size, $a_{\rm max}$, the increase in $a_{\rm align}$ corresponds to the increasing loss of grain alignment of grains having sizes smaller than $a_{\rm align}$ toward the filament's spine.

Figure \ref{fig:muscamodel} (lower panel) compares the polarization profile obtained from our modeling, $P_{\rm mod}$, given by Equation \ref{eq:Pmod2}, with the observational data for the case without RAT-D. Musca has low temperatures $<19\;\K$ and high densities of $10^2 -10^3\cm^{-3}$, therefore, RAT-D cannot occur. Models with only RAT-A can very well reproduce the data. We note that the B-field tangling is expected to have a minor effect on the polarization hole in Musca because the values of polarization angle dispersion function $\S$ are small ($<10^\circ$) and do not change considerably across the filament (Figure \ref{fig:musca_profiles} b). In addition, there is little dependence of $P$ on $\S$ with $\eta = 0.03$.

Figure \ref{fig:muscamodel} shows model predictions of polarization fraction with different values of maximum grain sizes varying from 0.25 to \mbox{0.5 $\mum$}. Dust grains can be aligned between $a_{\rm align}$ and $a_{\rm max}$; therefore, higher $a_{\rm max}$ produces a higher polarization fraction. Moreover, as $a_{\rm max}$ increases, the size distribution of aligned grains becomes broader, narrowing the difference between the polarization fraction between the outer and inner of Musca. As shown, our polarization models with $a_{\rm max}=0.25-0.35\,\mum$ can successfully reproduce the strong decrease in the polarization fraction toward the filament's spine. Therefore, the loss of grain alignment by RATs (i.e., increasing $a_{\rm align}$) toward the denser and colder region of the filament (upper panel) is the main origin of the polarization hole observed in Musca. 


\subsubsection{OMC-1}
For OMC-1 that has the structure and physical conditions varying significantly, such as between the east and west (with/without \ion{H}{2} region) and the presence/absence of an embedded protostar within the filament, we will perform pixel-by-pixel modeling for the whole region to obtain maps of polarization fraction to compare with the observational data.

\label{subsec:orionmodel}

\begin{figure*}[!htb]
\centering
\includegraphics[trim=0.cm 0cm 0.cm 0cm,clip,width=5.75cm]{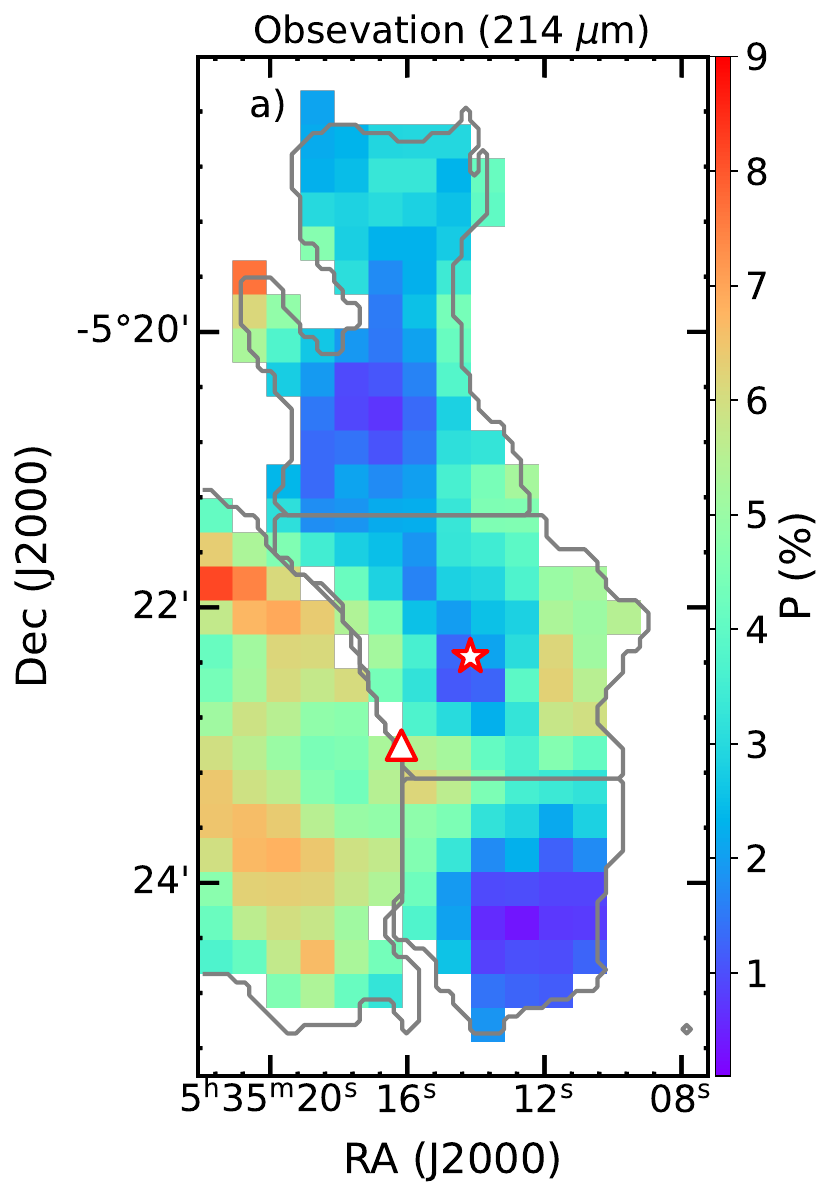}
\includegraphics[trim=0cm 0cm 0.cm 0cm,clip,width=5.5cm]{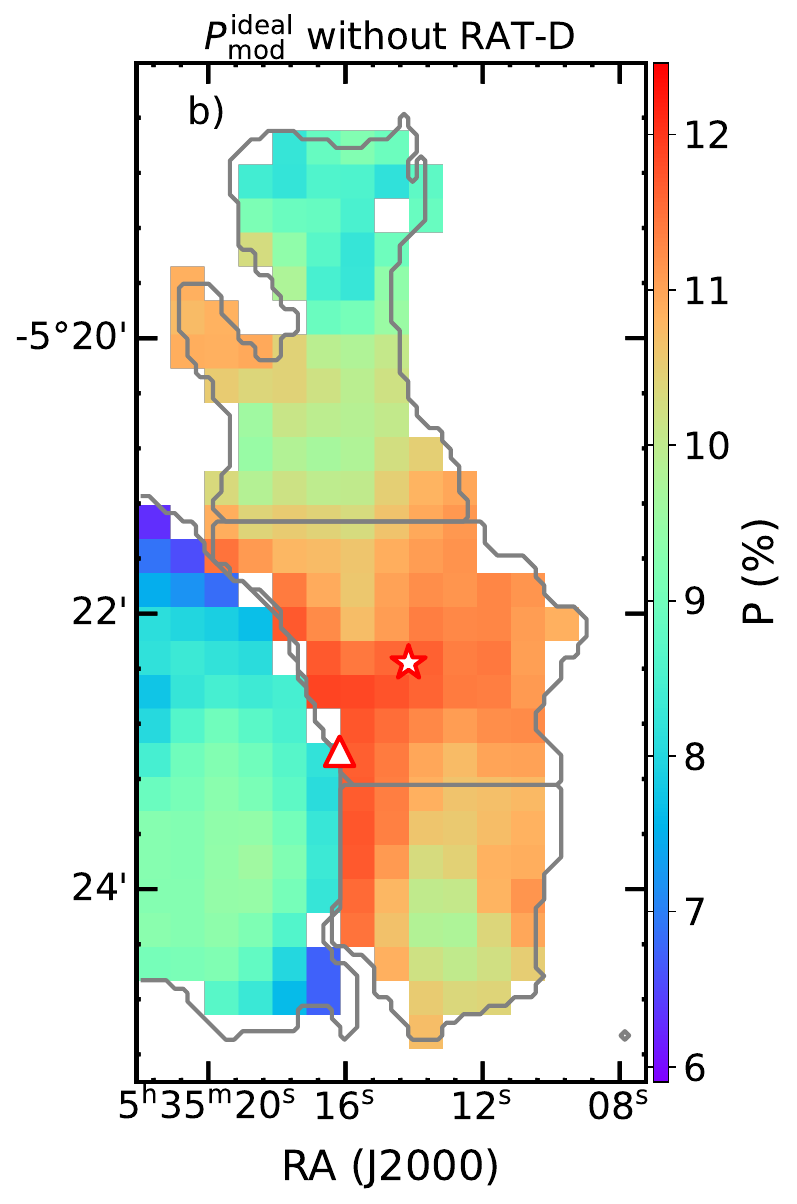}
\includegraphics[trim=0cm 0cm 0.cm 0cm,clip,width=5.5cm]{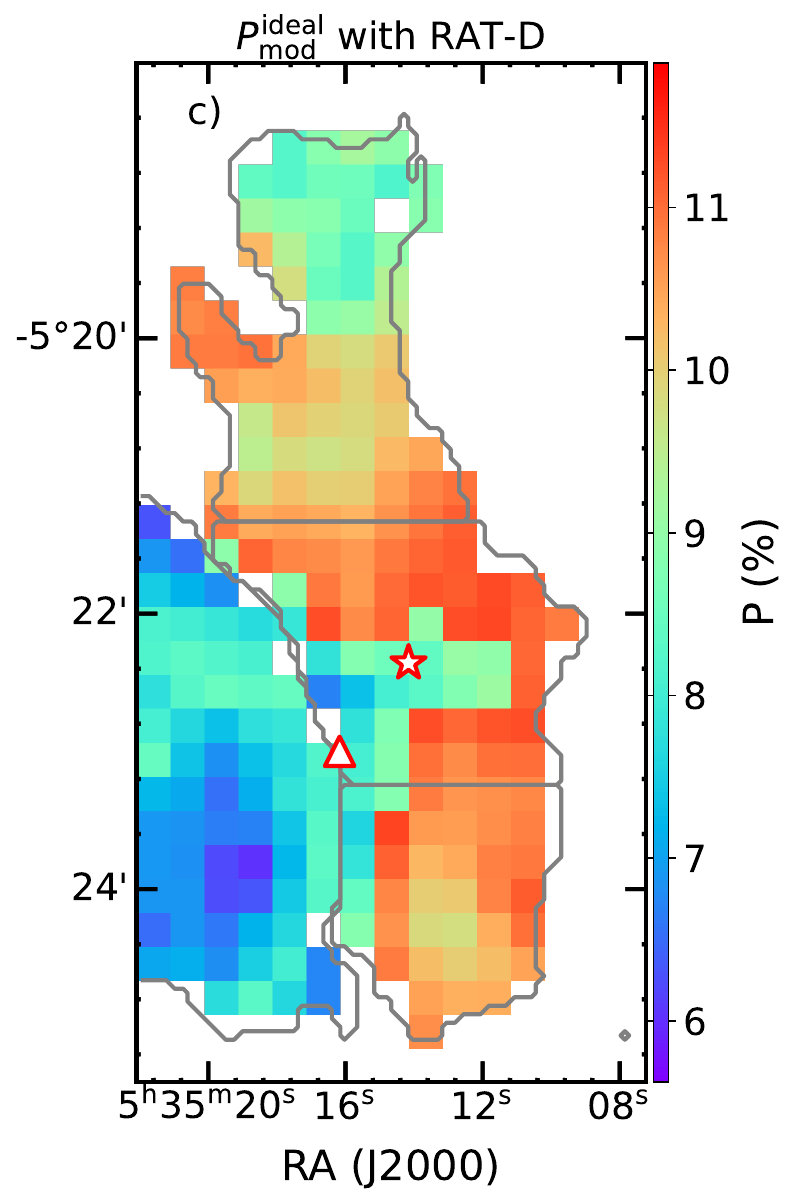}
\includegraphics[trim=0.cm 0cm 0.cm 0cm,clip,width=5.9cm]{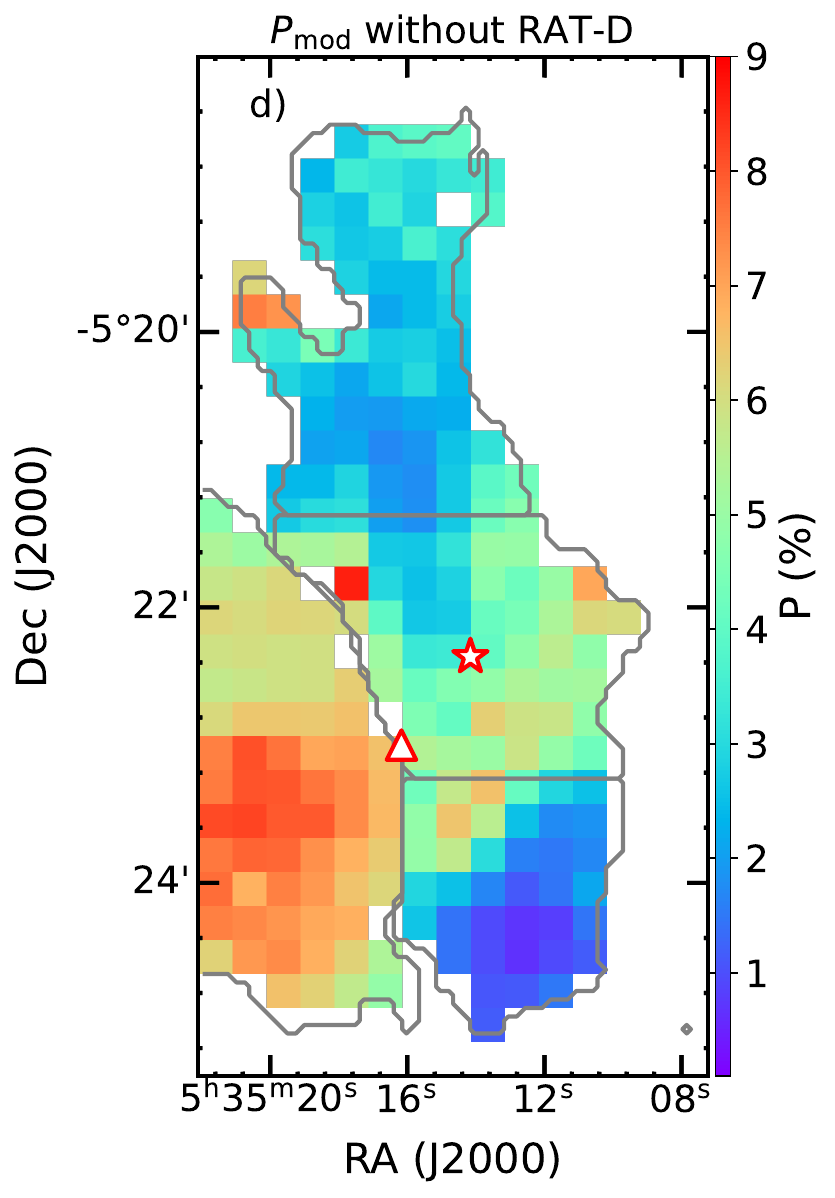}
\includegraphics[trim=0cm 0cm 0.cm 0cm,clip,width=5.5cm]{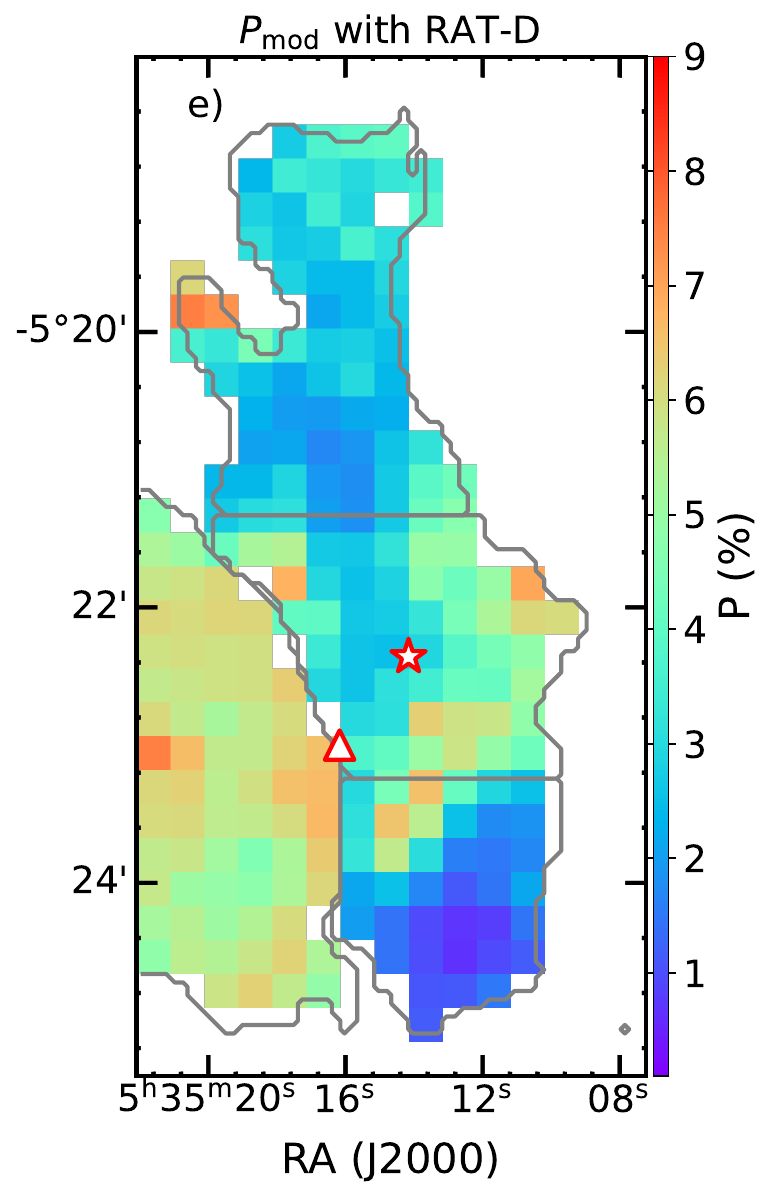}
\caption{OMC-1. The polarization fraction maps, (a) from observation and (b)-(e) from numerical modeling: (b) $P_{\rm mod}^{\rm ideal}$ without RAT-D, (c) $P_{\rm mod}^{\rm ideal}$ with RAT-D, (d) $P_{\rm mod}$ without RAT-D, and (e) $P_{\rm mod}$ with RAT-D. The gray contour corresponds with 4 sub-regions defined in Figure \ref{fig:orionmaps}. The star indicates the location of BN/KL, and the triangle shows the location of the Trapezium cluster.}\label{fig:orion_model_map} 
\end{figure*}

\begin{figure*}[!htb]
\centering
\includegraphics[trim=0.cm 0.cm 0.cm 0cm,clip,width=16cm]{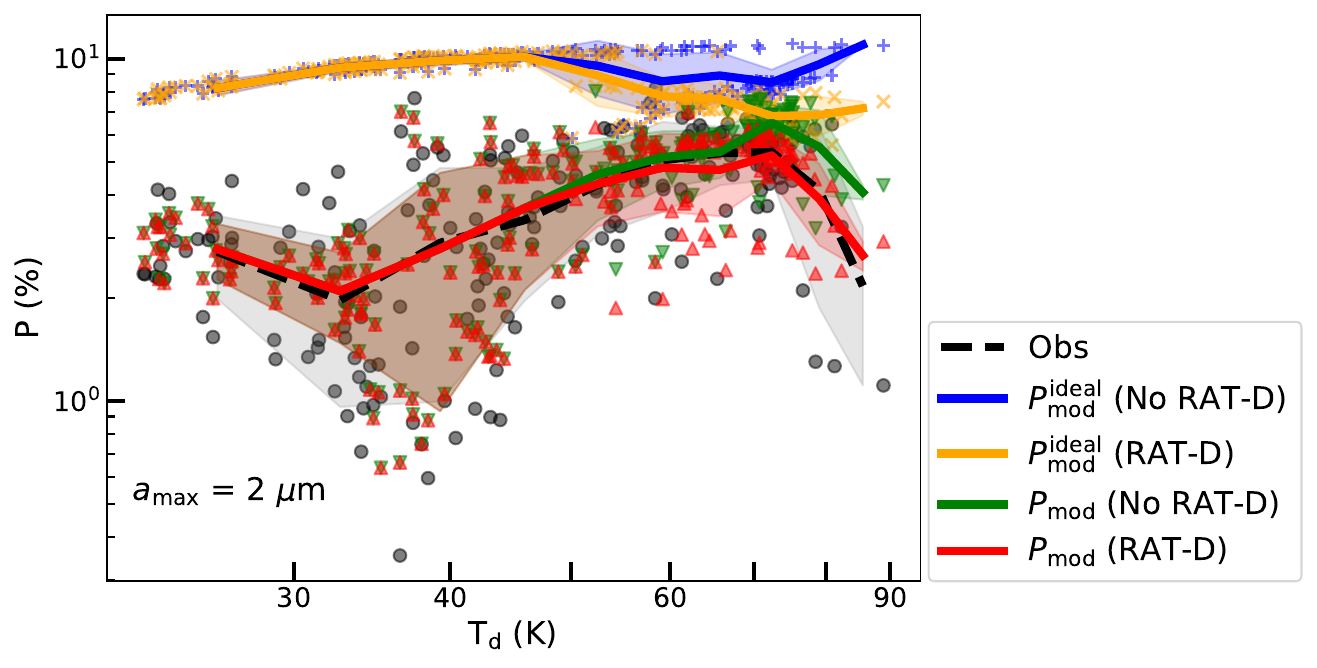}
\caption{OMC-1. Comparison of the polarization degree vs. dust temperature relation from our models (color lines) with observational data (black line). The shaded area represents the 1-$\sigma$ deviation of each bin of data points which are shown by points (black for observation data and color for model data). The ideal polarization model overestimates the observational data (blue and orange lines), but the realistic polarization model (red line) considering RAT-A, RAT-D, and B-field tangling ($P_{\rm mod}$) can best reproduce the observed variation of $P-T_{\rm d}$.}\label{fig:ori_model_pvst} 
\end{figure*}

\begin{figure*}[!htb]
\centering
\includegraphics[trim=0.cm 0.cm 0.cm 0cm,clip,width=16cm]{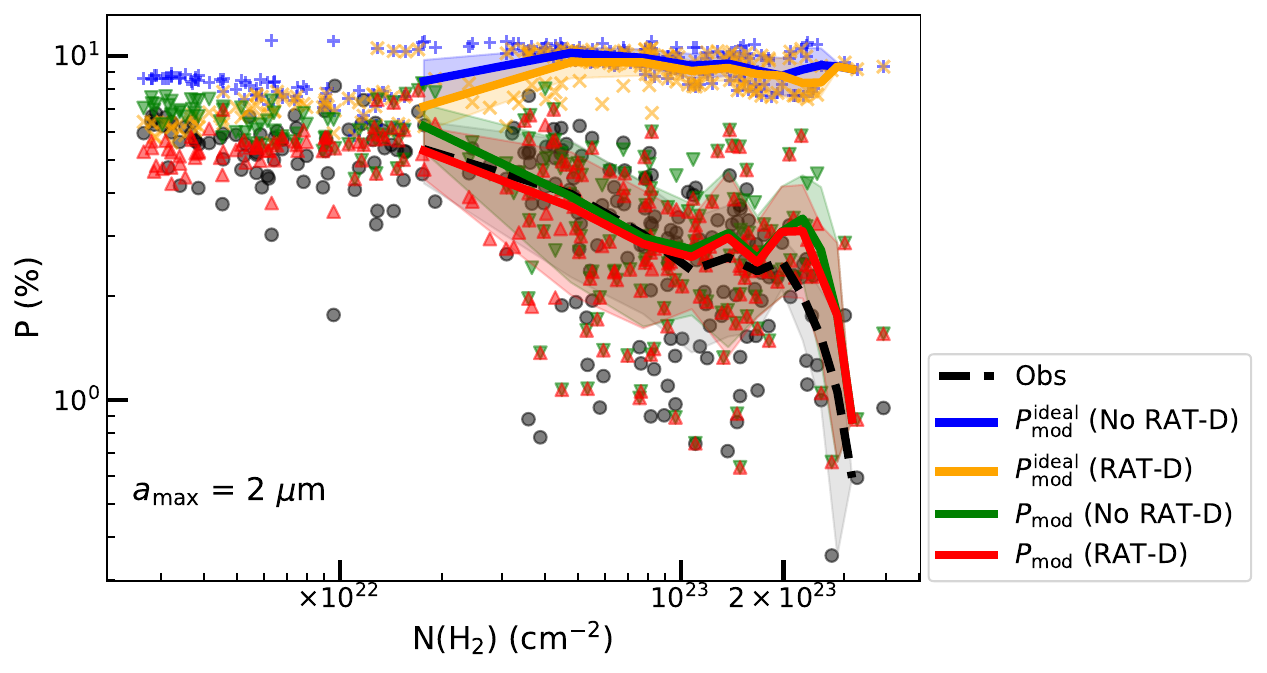}
\caption{OMC-1. Same as Figure \ref{fig:ori_model_pvst} but for $P-N({\rm H_2})$ relation. The ideal polarization model (blue and orange lines) cannot reproduce the data, but the realistic polarization model (red line) considering RAT-A, RAT-D, and B-field tangling ($P_{\rm mod}$) can best reproduce the observed polarization hole.}\label{fig:ori_model_pvsNH2} 
\end{figure*}

\begin{table*}[!htb]
\begin{center}
\caption{The comparisons between models and observations for different regions in OMC-1: North, BN-KL, South, \ion{H}{2}, and all. The statistical evaluation is shown by $\chi^2 = \frac{1}{N} \sum_{i=1}^{N} \frac{(P^{\rm i}_{\rm mod}-P^{\rm i}_{\rm obs})^2}{P^{\rm i}_{\rm obs}}$ with $N$ the number of pixels within each region. As shown, the values of $\chi^{2}$ are the highest for $P_{\rm mod}^{\rm ideal}$. $P_{\rm mod}$ with B-field tangling and RAT-D produces the smallest $\chi^{2}$, which implies that both grain disruption and B-field tangling play an important role in the observed polarization. Moreover, the model with the rotational disruption effect matches much better with observations in BN-KL and HII regions.}\label{tab:chisquare}
\begin{tabular}{c c c c c c c}
\hline
\hline
 Model &  & & $\chi^2$ & & \\
{} & North & BN-KL  & South & \ion{H}{2} & All \\\hline
$P_{\rm mod}^{\rm ideal}$ without RAT-D& 21.65 & 17.34  & 46.77 & 1.39 &  18.76 \\
$P_{\rm mod}^{\rm ideal}$ with RAT-D& 21.65 & 12.24  & 45.88 & 0.65 & 17.20 \\
$P_{\rm mod}$ without RAT-D& 0.32 & 0.83  & 0.30 & 0.32 & 0.43 \\
$P_{\rm mod}$ with RAT-D& 0.32 & 0.55  & 0.4 & 0.17 & 0.34 \\
\hline \hline
\end{tabular}
\end{center}
\end{table*}

For the modeling of the OMC-1 molecular cloud, we assume the mean radiation wavelength to be $\bar{\lambda} \simeq 0.53 \cm \K /T_{\star}$ \citep{hoang2021polhole}, which corresponds to $\bar{\lambda}= 0.3\,\mum$ for the presence of the nearby massive OB star cluster of $T_{\star}> 2 \times 10^4$ K. The pixel-by-pixel maps of the number density, $n_{\rm H_2}$, and dust temperature, $\Td$, are used for modeling grain alignment and polarization fraction. To derive the number density map, OMC-1 is assumed to have the same depth $d_{\rm OMC} = 0.15 \pc$, as obtained by \cite{chuss2019hawc+}. Therefore, the volume density is accordingly calculated by $n_{\rm H_2} = N({\rm H_2})/d_{\rm OMC}$. To reduce measurement noise and optimize model computation time, we increase the pixel size by a factor of $\sim$4 to be $14\arcsec$, almost equal to the spatial resolution of the input data maps.

Since grain growth is expected to be efficient in dense star-forming regions, we assume the maximum grain size of $a_{\rm max} = 2\,\mum$ for $N(\rm H_2) > 2 \times 10^{22}$ (cm$^{-2}$) in OMC-1. In the remaining regions of low density, such as the outer region of the filament structure, and the \ion{H}{2} region, we assume the maximum grain size of $a_{\rm max} = 0.2 \,\mum$, similar to that of the diffuse ISM. When modeling the RAT-D effect expected to be important in OMC-1 due to strong radiation fields, the tensile strength of dust grains is required. Because the tensile strength of dust grains depends on the grain internal structure (e.g., porosity) which varies with the grain size due to grain coagulation \citep{Tatsuuma19}, we assume $S_{\rm max} = 10^7 \erg \cm^{-3}$ for large grains having $a_{\rm max} = 2\,\mum$, and a higher tensile strength of $S_{\rm max} = 10^8 \erg \cm^{-3}$ for grains having $a_{\rm max} = 0.2\,\mum$. 

As shown in Figure \ref{fig:orionrelation} (c), the angle dispersion function in OMC-1 can go up to 50$^\circ$, and the slopes of $P-\S$ relation are 0.47--0.69 in the filament, which means that B-field tangling in OMC-1 is stronger than Musca and varies significantly across the object. Therefore, one cannot take the dust polarization results directly obtained by DustPOL-py  as in Musca. We use the polarization model, which takes into account the effect of B-field fluctuations given by Equation \ref{eq:Pmod2} for different subregions in OMC-1 with different $\eta$ (see Section \ref{sec:model}).

Figure \ref{fig:orion_model_map} (b) - (e) shows the polarization fraction maps obtained by using our numerical modeling. Figure \ref{fig:orion_model_map} (a) shows the observational data for comparison. We note that the pixel sizes in Figure \ref{fig:orion_model_map} now are $\sim$$14 \arcsec$. From Figure \ref{fig:orion_model_map} (b), one can see that $P_{\rm mod}^{\rm ideal}$ without RAT-D can reproduce a part of the South and North polarized holes; however, the model cannot reproduce the polarization hole in BN-KL. Figure \ref{fig:orion_model_map} (c) shows the result from $P_{\rm mod}^{\rm ideal}$ with RAT-D; the polarization fraction decreases in the BN-KL region, but $P$ from the model is significantly higher than the observed data. The models in Figure \ref{fig:orion_model_map} (d) and (e) incorporate B-field tangling effects, $P_{\rm mod}$ without/with RAT-D with exponents $\eta =0.61,0.48, 0.69,$ and $0.17$ for the North, BN-KL, South, and \ion{H}{2} regions, respectively. Here, the values of $\eta$ are chosen based on the slopes of the $P - \S$ relation in Figure \ref{fig:orionrelation} (c). It is evident that $P_{\rm mod}$ without RAT-D is limited to reproducing observations in the southern and northern areas. However, when RAT-D is incorporated, our model aligns more closely with observations, including those in the BN-KL and \ion{H}{2} regions. In the ensuing analysis, we conducted a thorough comparison utilizing two distinct diagnostic methods for the $P-T_{\rm d}$ relationship (which directly reflect the impact of alignment and disruption) and $P-N({\rm H_2})$ (which directly reflects the gas damping effect).

Figure \ref{fig:ori_model_pvst} shows the comparison of the $P-T_{\rm d}$ relation produced by our model and from observations for the whole OMC-1. There is a noticeable correlation between $P$ and $\Td$ at low temperatures ($< 70 \K$), while an anti-correlation is observed at high temperatures ($> 70 \K$). It is clear that models that do not take into account the fluctuations of B-fields fail to reproduce the observations, while those incorporating both the RAT-D mechanism and B-field fluctuations provide the most satisfactory explanation for these two trends compared to other models.

Figures \ref{fig:ori_model_pvst_regions} and \ref{fig:ori_model_psvnh2regions} show the comparisons for the individual regions, including the North, South, BK-KL, and \ion{H}{2} regions. In the regions with lower dust temperatures (North and South), models with only RAT-A and a combination with RAT-D are identical because the disruption effect does not occur. For the regions of high dust temperatures (BN-KL and \ion{H}{2}), models with RAT-D and B-field fluctuations best reproduce the decrease of polarization degree at $T_{\rm d}>50$ K. 

Figure \ref{fig:ori_model_pvsNH2} compares the $P-N({\rm H_2})$ relations of our models with observations. One can see that models without B-field tangling, $P_{\rm mod}^{\rm ideal}$, produce a rather high polarization and cannot reproduce the polarization hole. On the other hand, models with the B-field tangling, $P_{\rm mod}$, better reproduce the data. Therefore, the main cause of the depolarization in $P-N({\rm H_2})$ is due to the B-field tangling, and the grain alignment is still efficient at large column density due to the effect of embedded radiation sources. This is different from the case of Musca, where the main cause of the polarization hole is the loss of grain alignment toward the denser and weaker radiation field.

We refer to Table \ref{tab:chisquare} for a quantitative $\chi^{2}$ comparison. As shown, the values of $\chi^{2}$ are larger for $P_{\rm mod}^{\rm ideal}$ and smaller for $P_{\rm mod}^{\rm ideal}$ with the RAT-D effect. It is improved for the BN-KN and \ion{H}{2} regions where RAT-D happens. $P_{\rm mod}$ with the B-field tangling produces the smaller $\chi^{2}$, which implies that both grain alignment and B-field tangling play a role in the observed polarization. Therefore, we see that the best model is $P_{\rm mod}$ with both RAT-D and B-field tangling taken into account.

\section{Discussions}
\label{sec:discuss}
In this section, we will discuss the evidence of grain alignment and disruption by RATs obtained from comparing modeling results with observational data for Musca and OMC-1.

\subsection{Evidence of RAT-A Mechanism}
\label{subsec:eviRATA}

\subsubsection{Decrease of polarization fraction with column density: Origin of polarization hole}

The anti-correlation between $P$ and $\NHt$ (or $Av$), aka. polarization hole, is usually found in various environments, from the diffuse ISM to molecular clouds, filaments, and star-formation regions (e.g., \citealp{pattle2019a}). Yet, the exact origin of the polarization hole remains unclear. The loss of grain alignment toward denser regions by RATs is a leading explanation for the polarization hole \citep{hoang2021polhole}. Qualitatively, according to the RAT-A theory, grains are only efficiently aligned with B-fields if the grain size is larger than the minimum alignment size of $a_{\rm align} \propto n^{2/7}_{\rm H} \Td^{-12/7}$ (see Equation \ref{eq:a_align}). The alignment size increases with increasing gas density and decreasing $T_{\rm d}$, leading to a narrower size distribution of aligned grains and the lower dust polarization fraction \citep{lee_2020}. 

In this paper, to quantitatively test the RAT-A theory, we analyzed the thermal dust polarization data toward one of the simplest filaments, Musca, that exhibited a clear polarization hole. We performed a detailed polarization modeling for Musca using the local volume density and radiation fields (dust temperatures) from observational data as the input parameters for the DustPOL-py  code. Our modeling results showed that the alignment size ($a_{\rm align}$) increases rapidly from the outer to inner filament when the density increases and dust temperature decreases (see upper panel of Figure \ref{fig:muscamodel}). As a result, the polarization degree ($P$) decreases rapidly toward the filament spine, successfully reproducing the polarization hole observed toward Musca (see the lower panel of Figure \ref{fig:muscamodel}).

\subsubsection{Increase of polarization fraction with dust temperature}

\begin{figure}[!htb]
\centering
\includegraphics[trim=0.cm 0.cm 1.5cm 0cm,clip,width=8cm]{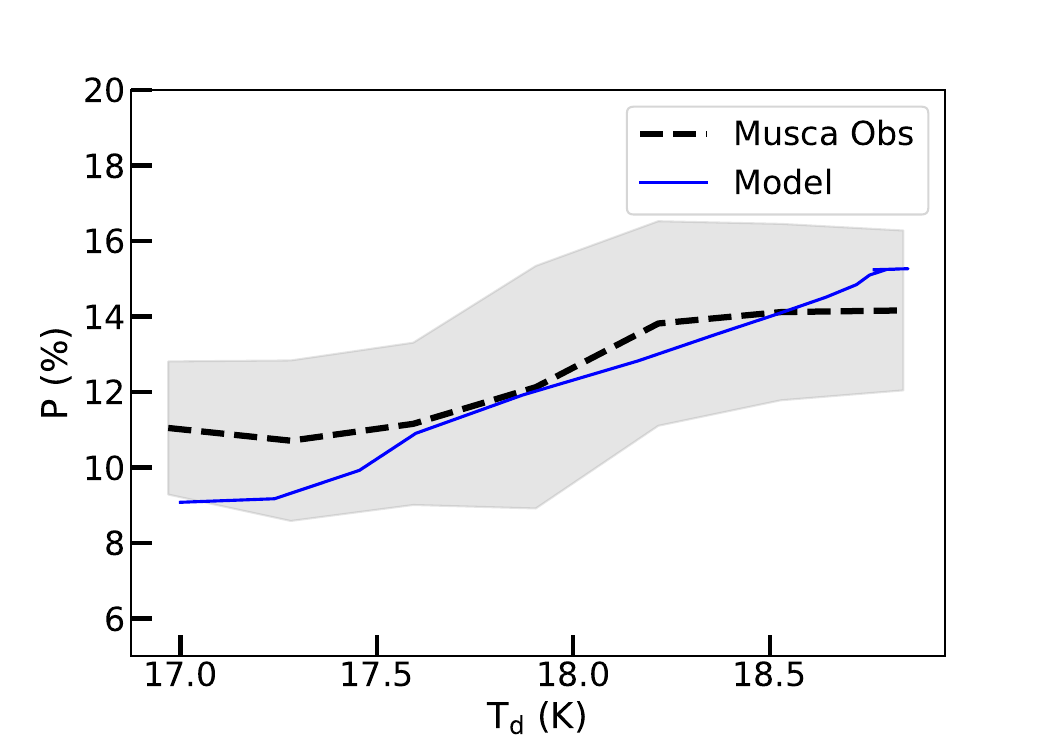}
\caption{Musca: Comparison of the polarization degree vs. dust temperature relation from our models (blue line) with observational data (black line). The shaded area represents the 1-$\sigma$ deviation of each bin of data points. The observed increase of $P$ with $T_{d}$ can be successfully reproduced by the polarization model based on the RAT-A theory.}\label{fig:musca_model_pvst} 
\end{figure}

The second prediction of the RAT-A mechanism is the increase of the polarization degree with increasing dust temperature because the alignment size decreases with $T_{\rm d}$ (see Equation \ref{eq:a_align}) and the size distribution of aligned grains becomes broader \citep{lee_2020}.

For both filaments, the polarization data indeed show the increase of $P$ with $\Td$ for low temperatures (see Section \ref{sec:Analysis}). However, for OMC-1 with strong radiation fields, the polarization degree reverses its trend at sufficiently high temperatures, which suggests the signature of RAT-D. Our detailed modeling for Musca and OMC-1 successfully reproduce the rising trend of $P-T_{\rm d}$ (see Figures \ref{fig:musca_model_pvst} and \ref{fig:ori_amax}).

\subsubsection{Role of B-field tangling in producing polarization hole: The anti-correlations of $P\;vs.\;\S$ and $P\;vs.\;N({\rm H_2})$}

Together with grain alignment and properties, polarization fraction, $P$, also depends on the fluctuations of B-fields along a light of sight and within the beam \citep{HoangTruong23}. The polarization angle in the POS is the average of the polarization patterns, which are aligned with the local B-fields along a LOS. Therefore, polarization angle dispersion function $\S$ is usually used to describe the B-field fluctuations.

In Musca, where the B-field fluctuations are small (see Figure \ref{fig:musca_profiles}, our polarization model based on the RAT paradigm could reproduce the observed polarization data (see Figure \ref{fig:muscamodel}). Therefore, the loss of RAT alignment is the main cause of the polarization hole, and the effect of B-field tangling is negligible. 

The situation becomes different for OMC-1, where B-field fluctuations are strong and spatially dependent (see Figure \ref{fig:orionrelation2}). Our detailed modeling results show that the model without B-field tangling, $P_{\rm mod}^{\rm ideal}$, produces too high polarization and cannot reproduce the data. Only the polarization models including B-field tangling effects, $P_{\rm mod}$, could reproduce the observational data (see Figure \ref{fig:ori_model_pvst}). This reveals the important role of B-field tangling in producing the polarization hole in OMC-1.


\subsection{Evidence of RAT-D Mechanism in OMC-1}
\label{subsec:eviRATD}
The correlation of $P$ and $\Td$ at low dust temperature and anti-correlation at high dust temperature is found in the OMC-1 filament. The decreasing trend at high dust temperature is the opposite of what would be expected from the RAT-A theory, but a signature of RAT-D.

To quantitatively study whether RAT-D could reproduce the declining trend in OMC-1, we performed detailed modeling of dust polarization using the maps of the gas volume density and dust temperature for each pixel obtained from observations. Our modeling results could reproduce the $P\;{\rm vs.}\;\Td$ relation (see Figure \ref{fig:ori_model_pvst}). 

We note that the B-field tangling ($P_{\rm mod}$) alone cannot reproduce the decrease of $P$ with $\Td$ at high temperatures. Indeed, as shown in Figure \ref{fig:orionrelation2}, the polarization angle dispersion function $\S$ tends to decrease or change slowly at high $\Td$. Our models that take into account both the B-field tangling and RAT-D shown in Section \ref{sec:model} could produce the observed $P-\Td$ relation at high $\Td$. Therefore, the RAT-D effect can explain the decreasing trend of $P-T_{\rm d}$ in OMC-1.

In previous simple modeling and data analyses using SOFIA/HAWC+, tentative evidence of RAT-D in some regions of OMC-1 was reported, e.g., in BN-KL region \citep{tram2021bnkl} and in Orion Bar \citep{LeGouellec2023}. It is worth noting that evidence of RAT-D was previously found in several star-forming regions: Auriga \citep{ngoc2021observations}, Oph A \citep{tram2021ophiuchi}, 30 Doradus \citep{tram2021doradus}, and M17 \citep{thuong22}.

\begin{figure*}[!htb]
\centering
\includegraphics[trim=0.5cm 0.2cm 1.cm 0.5cm,clip,width=8.5cm]{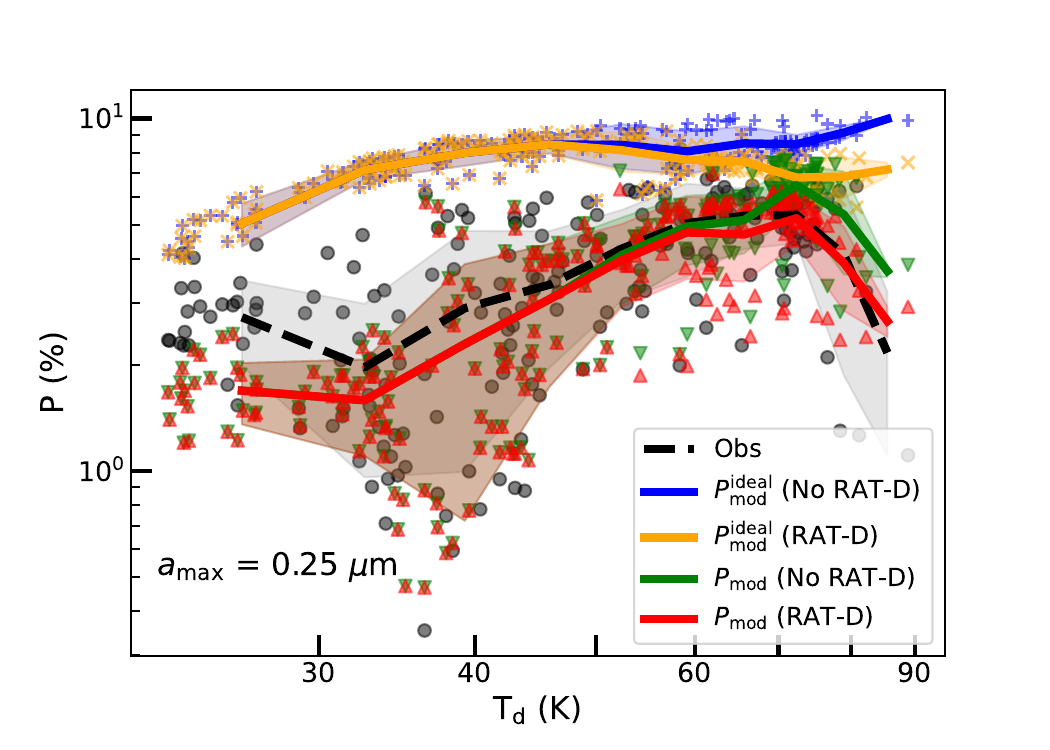}
\includegraphics[trim=0.5cm 0.2cm 1.cm 0.5cm,clip,width=8.5cm]{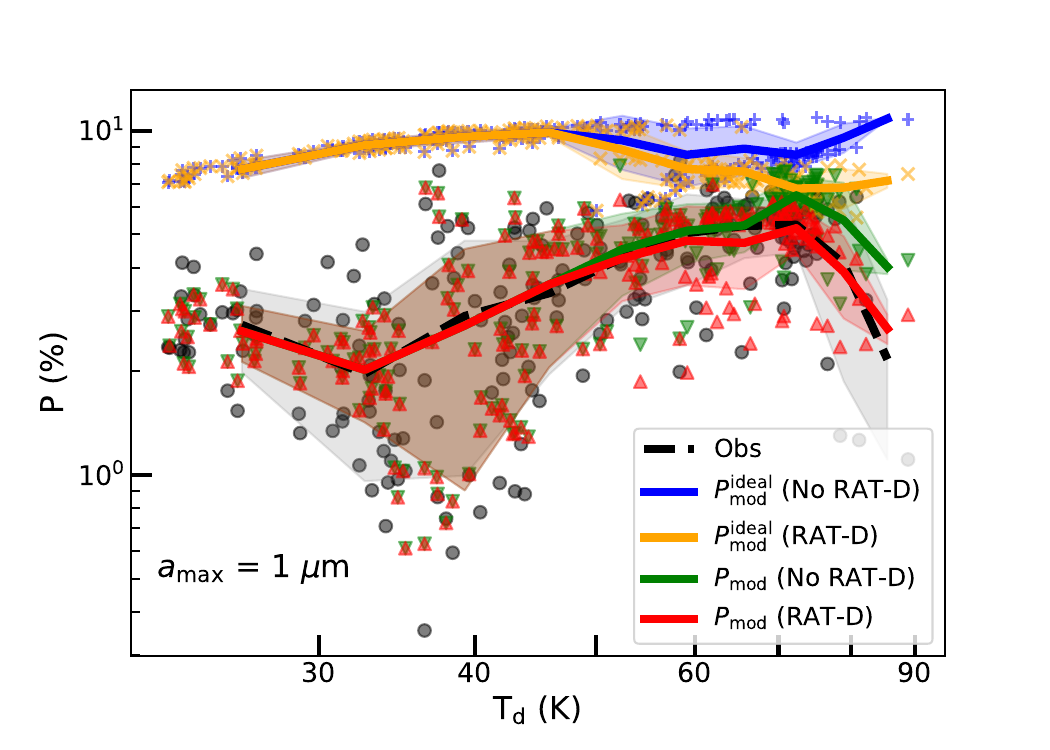}
\caption{OMC-1. Comparison of the polarization fraction between our models and observations for different initial maximum sizes of $a_{\rm max} =0.25\,\mum$ (left panel) and $a_{\rm max} =1\,\mum$ (right panel). The latter realistic models (red line) can reproduce better the data for the region of high density but low temperatures of $\Td<50\K$, revealing the significant grain growth in this dense region.}\label{fig:ori_amax} 
\end{figure*}

\subsection{Implication for Probing Grain Growth}
Observational studies suggest that the upper limit of the grain size distribution in the ISM is $\sim$$0.25 \,\mum$ \citep{mathis1977size}. However, grain growth is expected to occur within dense molecular clouds or filaments due to gas accretion and grain coagulation. \citet{Vaillancourt:2020ch} reveal grain growth in dense clouds using observations of starlight polarization combined with numerical modeling using the RAT theory. Using polarization data, \citet{Ngoc23} gives evidence that grain growth is moderate with $a_{max}>0.3\,\mum$ in the G11 filament when the volume density is higher than $10^4 \cm^{-3}$.

In this study, our numerical modeling for Musca shown in Figure \ref{fig:musca_model_pvst}  reveals that the maximum grain size is $\sim 0.3-0.35\,\mum$. This implies grain growth is effective but not significant in this not-dense filament, with the highest density of $n_{\H}\sim 10^{3}\cm^{-3}$.
On the other hand, the density of the OMC-1 filament varies significantly and can reach a high value of $10^4 -10^6\;\cm^{-3}$. To probe grain growth in this dense filament, we varied $a_{\rm max}$ from $0.2$ to $2\;\mum$ and found that the model with $a_{\rm max} > 1\; \mum$ can well reproduce the $P-\Td$ relation, in particular the low $\Td$ region in the lower dust temperature, $\Td < 50 \K$ (see Figure \ref{fig:ori_amax} left). This suggests that grain growth is significant in the dense regions of OMC-1.





\subsection{Limitations of Our Modeling and Future Works}
\label{subsec:limit}
In this paper, we have used the hybrid approach to perform modeling of thermal dust polarization and test the RAT paradigm. We first used the DustPOL-py  code to calculate the polarization fraction predicted by the RAT theory for the ideal model with B-fields in the POS. To account for the effects of the inclination and the tangling of B-fields, we then incorporated a coefficient $\Phi$ to describe the inclination angle and a depolarization factor $\S^{-\eta}$ to describe the effect of B-field fluctuations (Equation \ref{eq:Pmod2}). The final realistic dust polarization model is based on the optically thin assumption, which is successfully tested with synthetic observations of MHD simulations using POLARIS in \cite{HoangTruong23}. Although our numerical modeling can successfully reproduce the observational data and support the RAT paradigm, there are several limitations. First, we have assumed that grains are aligned via the MRAT mechanism with $f_{\rm max}=1$, which is expected for grains with embedded iron clusters. Second, the coefficient $\Phi$ obtained by fitting the polarization model to the maximum observed polarization is assumed to be constant across the filament. Such an assumption is likely valid for Musca which has well-ordered B-fields and a small slope $\eta\sim 0.03$. However, it may not be realistic for a complex star-forming region like OMC-1, which has strong B-field fluctuations, and the slope $\eta$ is found to vary considerably from region to region in OMC-1. 


In our follow-up studies, we will perform synthetic dust polarization using MHD simulations with our updated POLARIS code \citep{giang21POLARIS}, which includes both the RAT paradigm and the radiative transfer process. A detailed comparison of synthetic simulations with observational data would help disentangle the different origins of the polarization hole in filaments, including grain alignment loss, disruption, and B-field tangling.  


\section{Conclusions} \label{sec:conclusions}
In this study, we utilize dust polarization data observed toward Musca and Orion to test the grain alignment and rotational disruption mechanisms by radiative torques (i.e., RAT paradigm). We investigate the relation of the polarization degree, $P$, with gas column density, $N(\rm H_2)$, and dust temperature, $\Td$, and compare them with theoretical predictions using the RAT paradigm. Additionally, the variation of $P$ with polarization angle function, $\S$, is explored to study the effects of B-field fluctuations. Our results are as follows:

\begin{enumerate}
\item Observational data show the existence of a polarization hole, i.e., the decrease of the polarization degree with increasing the gas column density in both Musca and OMC-1. The analysis of the observed polarization angle dispersion reveals that B-fields fluctuations are rather weak in Musca but strong in OMC-1.

\item For Musca with weak B-field fluctuations, our detailed modeling of dust polarization using the RAT theory with small B-field tangling effects succeeds in reproducing the polarization hole, suggesting the loss of RAT alignment as the main cause of the polarization hole in this filament. Moreover, our modeling results show that grain growth is effective but not significant in Musca.

\item For the high-mass star-forming region, OMC-1, we find that, in general, polarization fraction ($P$) and polarization angle dispersion function ($\S$) are anti-correlated but $N(\rm H_2)$ and $\S$ are correlated, especially at low temperatures. This trend suggests that the B-field fluctuations in dense regions dominantly contribute to the depolarization effect.

\item Observational data exhibited a general correlation of $P$ with $\Td$ at low temperatures, such as in the northern and southern. In the Orion BN-KL region of high temperatures, $P$ reverses the trend and decreases with $\Td$. 

\item
Using the pixel-by-pixel polarization modeling method with local physical parameters (gas density and radiation fields) inferred from observations as the input, we demonstrate that the observed $P-T_{\rm d}$ and $P-N({\rm H_2})$ (polarization hole) relations can be simultaneously reproduced by our polarization models incorporating both RAT-A and RAT-D, and the contribution of B-field fluctuations. The decreasing trend of $P-T_{\rm d}$ when taking into account the RAT-D effect from models is in agreement with observations from the BN-KL region.

\item The successful testing of the RAT paradigm through the combination of pixel-by-pixel polarization modeling using DustPOL-py  with observational data to two different filaments of quiescent one (Musca) and an active star-forming region (OMC-1) opens the avenue for testing the RAT paradigm and constraining dust properties in other astrophysical environments.


\end{enumerate}

\software{Astropy \citep{robitaille2013astropy}, APLpy \citep{APLpy2012,Aplpy2019}}
\facility{Stratospheric Observatory for Infrared Astronomy (SOFIA), {\it Planck}}

\section{Acknowledgement}
The authors greatly appreciate the anonymous reviewer for carefully reading the manuscript and providing valuable comments and suggestions. The authors sincerely thank Dr. Hyeseung Lee for her help with DustPOL-py in the initial stage of this work. We thank Dr. Lars Bonne for discussions regarding the Planck data, Dr. Joseph Michail and Prof. David Chuss for sharing the thermal dust polarization, dust temperature, and gas column density of OMC-1. This research was funded by Vingroup Innovation Foundation under project code VINIF.2023.DA.057. T.H. is supported by a grant from the Simons Foundation to IFIRSE, ICISE (916424, N.H.) N.B.N. was funded by the Master, Ph.D. Scholarship Programme of Vingroup Innovation Foundation (VINIF), code VINIF.2023.TS.077. This research is based on observations made with the NASA/DLR Stratospheric Observatory for Infrared Astronomy (SOFIA). SOFIA is jointly operated by the Universities Space Research Association, Inc. (USRA), under NASA contract NNA17BF53C, and the Deutsches SOFIA Institut (DSI) under DLR contract 50 OK 0901 to the University of Stuttgart. Based on observations obtained with Planck (\url{http://www.esa.int/Planck}), an ESA science mission with instruments and contributions directly funded by ESA Member States, NASA, and Canada.

\bibliography{filaments}{}

\bibliographystyle{aasjournal}

\appendix
\counterwithin{figure}{section}
\counterwithin{table}{section}

\section{Appendix}

\subsection{Physical Modeling Method}
\label{appendix:model}

The physical modeling of polarized thermal dust emission based on the RAT theory is described in detail in \cite{lee_2020} and \cite{tram2021ophiuchi}. Here, for reference, we summarize the main points for reference.

\subsubsection{Grain Alignment by Radiative Torques}
\label{subsubsec:align}
We consider a radiation field with a total energy density 
$u_{\rm rad}=\int u_{\lambda}d\lambda\; (\erg \cm^{-3})$. $U = u_{\rm rad}/u_{\rm ISRF}$ is radiation strength, where $u_{\rm ISRF} = 8.64 \times 10^{-13}$ (erg cm$^{-3}$) is the radiation energy density of the interstellar radiation field (ISRF) in the solar neighborhood \citep{mathis1983interstellar}. One important effect of RATs is to spin up dust grains of irregular shapes to suprathermal rotation \citep{draine1997radiative}. On the other hand, gas collisions and IR re-emission spin down grains. The balance between spin up and spin down establishes the maximum angular velocity of grains achieved by RATs.

The maximum angular velocity in units of thermal angular velocity spun up by RATs (e.g., \citealt{hoang2021polhole})
\bea
\frac{\omega_{\rm RAT}}{\omega_{\rm T}} &\simeq& 48.7\hat{\rho}a_{-5}^{3.2}U\left(\frac{\gamma}{0.1}\right) \left(\frac{30 \cm^{-3}}{n_{\rm H}}\right)\left(\frac{\bar{\lambda}}{1.2\;\mum}\right)\nonumber\\
&&\times\left(\frac{100\;\K}{T_{\rm gas}}\right)(1+F_{\rm IR})\label{eq:angvelo_rat}
\ena
where $a$ is the grain size, $a_{-5} = a/(10^{-5} \cm )$, $\rho$ is grain mass density, $\hat{\rho} = \rho/(3 \g \cm^{-3)}$, and $\bar{\lambda}=1.2\,\mum$ is a mean wavelength of the ISRF \citep{draine1997radiative}.  $F_{\rm IR}$ is a dimensionless coefficient of rotational damping by IR emission and is calculated by:
\bea
F_{\rm IR} \simeq 0.91\left(\frac{U^{2/3}}{a_{-5}}\right)\left(\frac{30 \cm^{-3}}{n_{\rm H}}\right)\left(\frac{100\;\K}{T_{\rm gas}}\right)^{1/2}.
\ena

In the unified alignment theory based on MRAT \citep{hoang2016unified}, the joint action of suprathermal rotation by RATs and enhanced paramagnetic relaxation produces efficient alignment of dust grains. The minimum size for grain alignment, aka alignment size, $a_{\rm align}$, is defined by the grain size at which $\omega/\omega_{T}=3$ \citep{hoang2008radiative}.

\subsubsection{Grain size distribution and RAT-D}
\label{subsubsec:ratd}

The tensile stress of a spinning grain having angular velocity $\omega$ is calculated by $S = \rho \omega^2 a^{2/4}$ where $\rho$ is the mass density of the grain. For large grains in the strong radiation field, the angular velocity by RAT can be sufficiently large such that $\S$ exceeds the maximum tensile strength $S_{\rm max}$ of the grain material, resulting in rotational disruption of the grain into small fragments \citep{hoang2019ratdnatastro}. The critical angular velocity for the disruption is given by

\bea
\omega_{\rm crit} = \frac{2}{a}\left(\frac{S_{\rm max}}{\rho}\right) \simeq \frac{3.6 \times 10^8}{a_{-5}}S_{\rm max,7}^{1/2}\hat{\rho}^{-1/2},
\ena
with, $S_{\rm max,7} = S_{\rm max}/(10^7 \erg \cm^{-3})$.

We can obtain the critical size $a_{\rm disr}$ above which grains are disrupted following this equation:
\bea
\left(\frac{a_{\rm disr}}{0.1\, \mum}\right)^{2.7} \simeq 5.1\gamma_{0.1}^{-1}U^{-1/3}\bar{\lambda}_{0.5}^{1.7}S_{\rm max,7}^{1/2},\label{eq:adisr}
\ena
with, $\bar{\lambda}_{0.5} = \bar{\lambda}/(0.5 \,\mum)$, and strong fields of $U \gg 1$ \citep{hoang2019ratdnatastro}. 

The maximum size that grains are still disrupted by RAT-D is given by \cite{hoang2019ratdnatastro}:
\bea
a_{\rm disr, max} \simeq 1.7\gamma\bar{\lambda}_{0.5}\left(\frac{U}{\bar{n}\bar{T}_{\gas}^{1/2}}\right)^{1/2}\left(\frac{1}{1+F_{\rm IR}}\right),
\ena
where $\bar{n} = n_{\rm H}/(30 \cm^{-3})$ and $\bar{T} = T_{\rm gas}/(100\;\K)$.

Due to the RATD mechanism, dust grains of size between $a_{\rm disr}-a_{\rm disr,max}$ are converted into smaller size $a<a_{\rm disr}$, such that the size distribution of the original dust grains is modified. In this work, we adopt a power-law grain size distribution for both the original large grains and the smaller grains produced by disruption by a power-low index $\beta$ \citep{mathis1977size}:
\bea
\frac{1}{n_{\rm H}}\frac{dn_{\rm sil,car}}{da} = C_{\rm sil,car}a^{\beta} (a_{\rm min} \leq a \leq a_{\rm max}),
\ena

where $dn_{j}$ is the number density of grains of material $j=sil,carb$ between $a$ and $a+da$, $C_{\rm sil}$ and $C_{\rm car}$ are the normalization constants for silicate and carbonaceous grains, respectively. The smallest grain size is chosen as $a_{\rm min}=10 \Angstrom$, while the $a_{\rm max}$ is constrained by the RAT-D mechanism.

\subsection{Musca}\label{appendix:muscaradfil}
\begin{figure*}[!htb]
\centering
\includegraphics[trim=0.5cm 0.0cm 0.5cm 0.5cm,clip,width=5.1 cm]{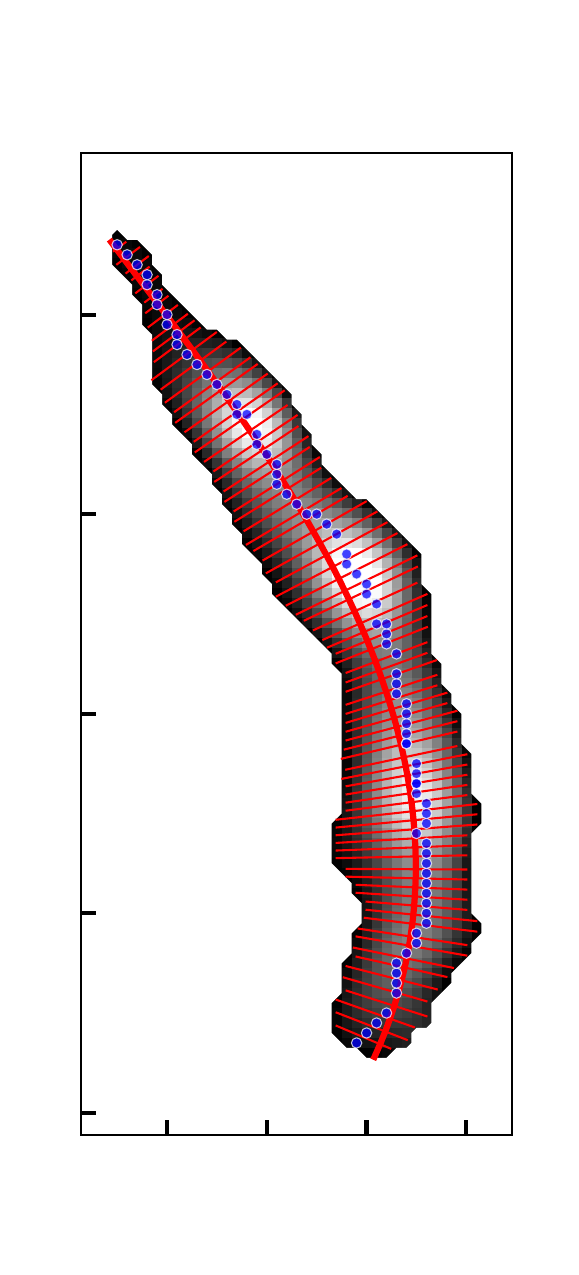}
\includegraphics[trim=0.4cm 0.0cm 0.0cm 1.cm,clip,width=12. cm]{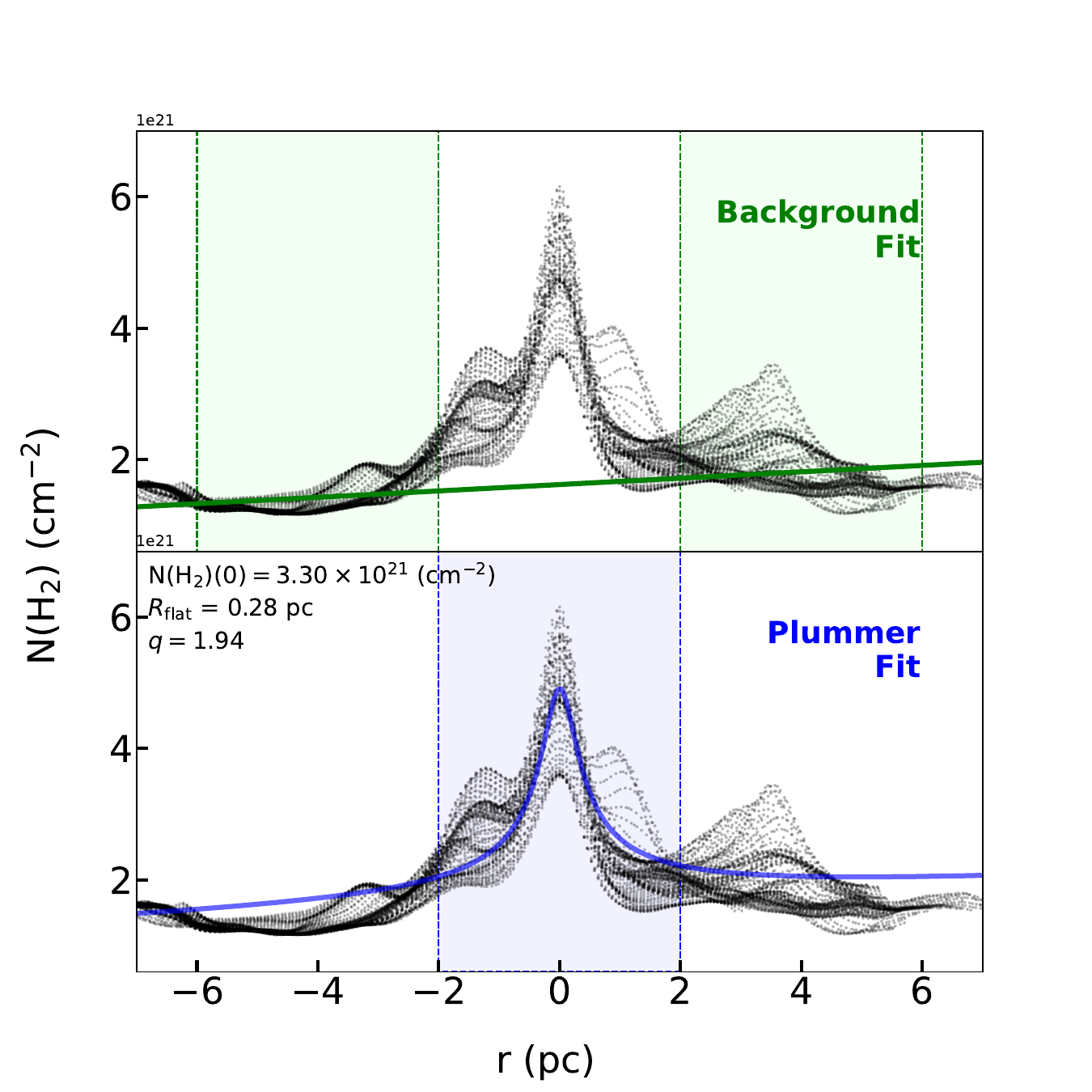}
\caption{Musca. Left: Column density map is used to build the radial profile of Musca by sampling radial cuts (red thin lines) with a step of 0.1 pc along the spine (thick red line). The spine is determined by {\it Radfil} using the provided mask using the B-spline smoothness algorithm \citep{zucker2018radfil}. The radial distance is defined as the projected distance from the considered pixel to the peak emission (blue dot) on a given cut. Right: Background fit (upper) and Plummer-like fit (lower). The results of the fit following Equation \ref{eq:plummer} are shown in the inserts. Green and blue shaded areas are the regions used for the background and Plummer-like fit, respectively.}
\label{fig:musca_radfil} 
\end{figure*}

We build the density profile and the spine of Musca using a Python package \textit{RadFil} applied for a column density map \citep{zucker2018radfil}. The obtained spine is shown by the thick red line over-plotted on the density map (grayscale background) in Figure \ref{fig:musca_radfil} (left). Subsequently, \textit{RadFil} generates radial cuts (red thin lines) perpendicular to the spine; from that, we have the radial profiles shown in black dotted curves in Figure \ref{fig:musca_radfil} (right). We estimated the background by subtracting a first-order polynomial from all profiles. The background fitting is performed within the radii \mbox{$2<|r|<6$ pc} (green area in the upper panel of Figure \ref{fig:musca_radfil}) right. Then, the fitted profile with a Plummer-like function again using \textit{RadFil} is shown by the thick curve in Figure \ref{fig:musca_radfil} (lower right).

\subsection{OMC-1}
\begin{figure*}[!htb]
\centering
\includegraphics[trim=0cm 0.cm 0.0cm 0cm,clip,width=13.5cm]{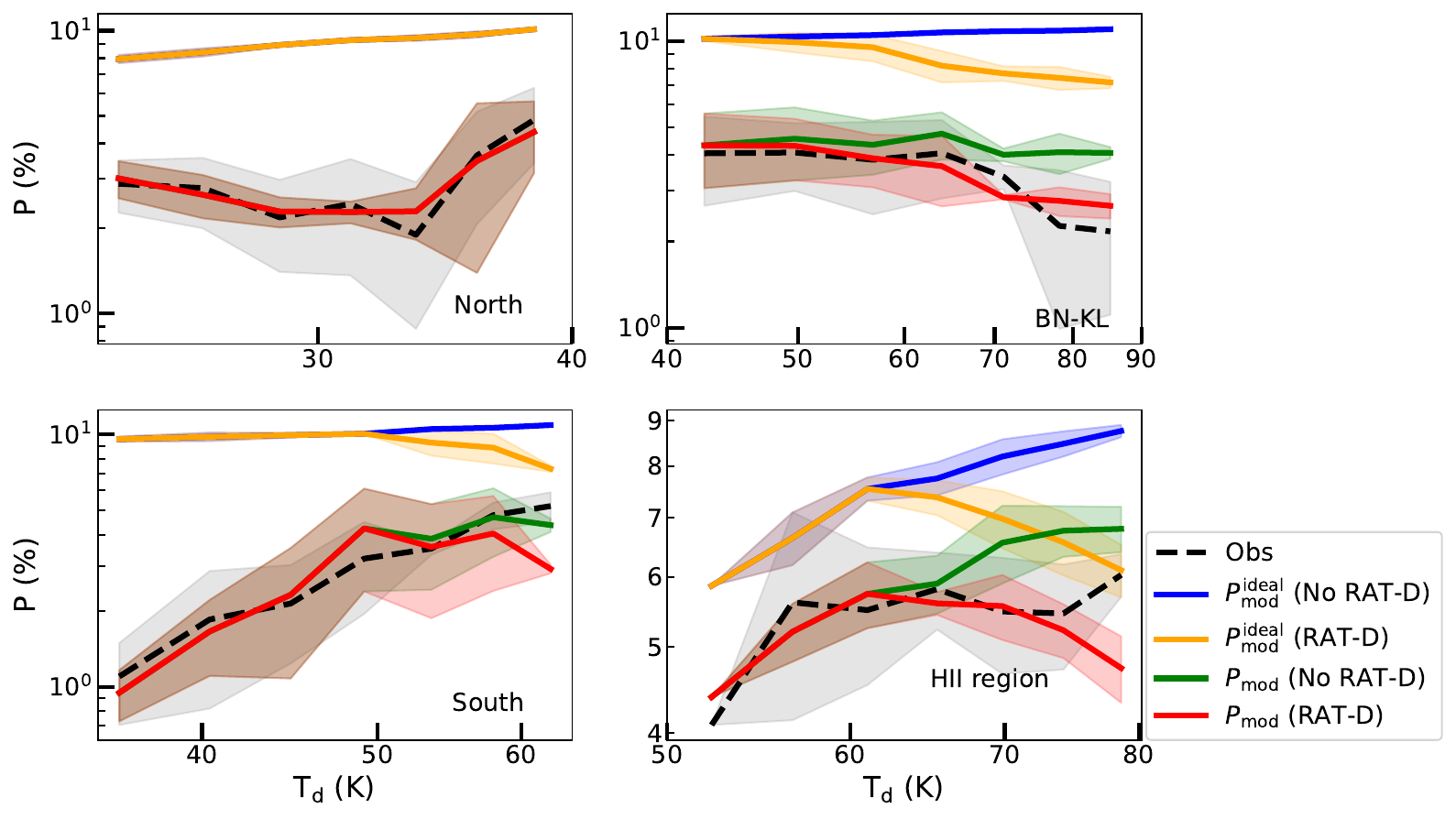}
\caption{OMC-1. Comparison of the polarization degree vs. dust temperature relation from our models (color lines) with observational data (dashed line). The shaded area represents the 1-$\sigma$ deviation of each bin of data points.}\label{fig:ori_model_pvst_regions} 
\end{figure*}

\begin{figure*}[!htb]
\centering
\includegraphics[trim=0cm 0.cm 0.0cm 0cm,clip,width=13.5cm]{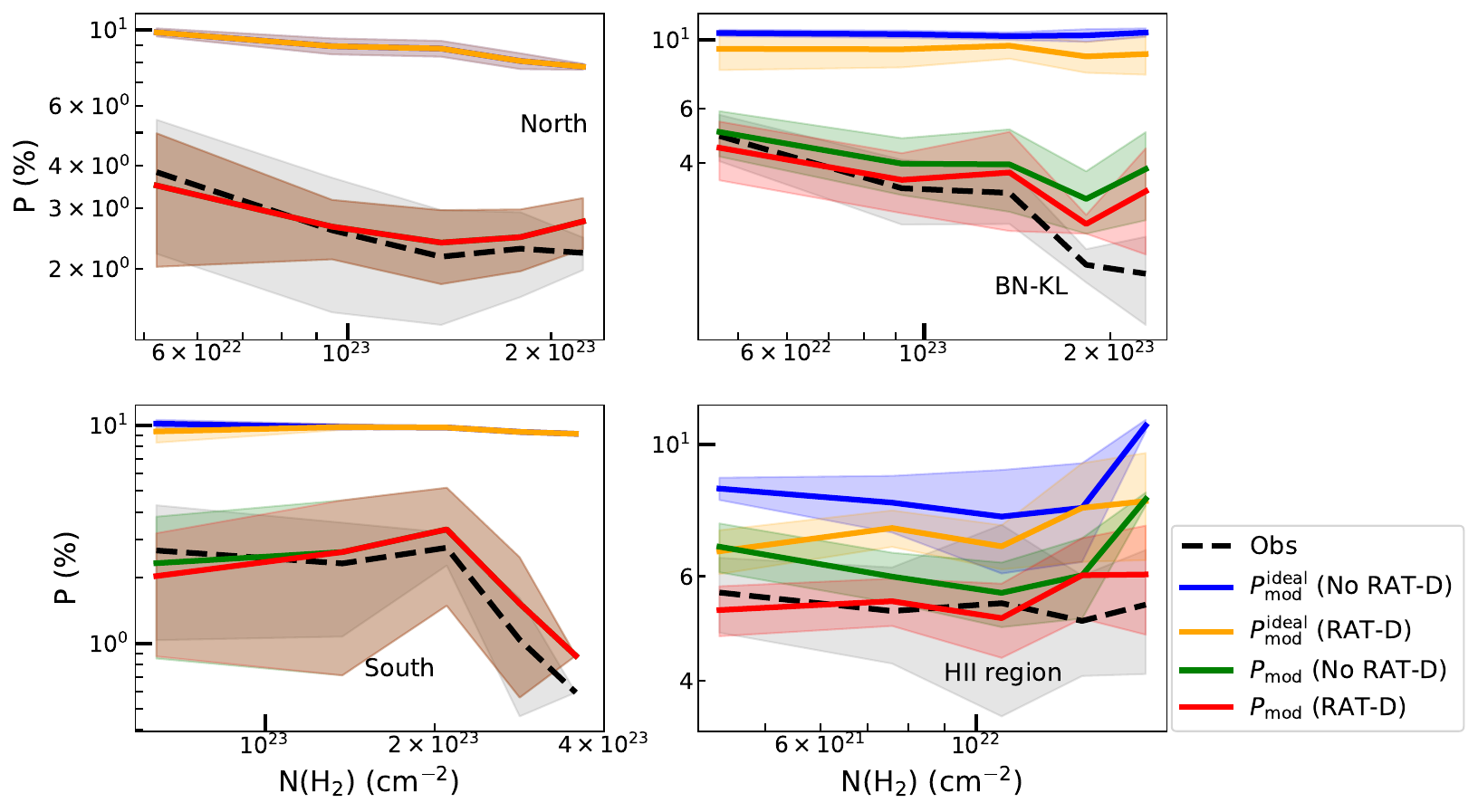}
\caption{OMC-1. Comparison of the polarization degree vs. column density relation from our models (color lines) with observational data (dashed line). The shaded area represents the 1-$\sigma$ deviation of each bin of data points.}\label{fig:ori_model_psvnh2regions} 
\end{figure*}

Figure \ref{fig:ori_model_pvst_regions} and Figure \ref{fig:ori_model_psvnh2regions} show in detail the comparison of our physical modeling results for $P-T_{\rm d}$ and $P-N_{\rm H_2}$, respectively, with the observational data for different regions in OMC-1, including North, South, BN-KL. The realistic polarization model with RAT-D best reproduces the observational data.

\end{document}